\documentclass[12pt]{article}
\usepackage{pgfplots}
\usepackage{graphicx}
\usepackage{epstopdf}
\usepackage{amsthm}
\usepackage{amsmath}
\usepackage{amssymb}
\usepackage{amstext}
\usepackage{amsfonts}
\usepackage{mathrsfs}
\usepackage{enumerate}
\usepackage{footnote}
\makesavenoteenv{tabular}
\makesavenoteenv{table}
\usepackage[onehalfspacing]{setspace}
\usepackage[shortlabels]{enumitem}
\usepackage{mathabx}
\usepackage{mathtools}

\usepackage{multirow}
\usepackage[title]{appendix}
\usepackage{xcolor}
\usepackage{setspace}
\usepackage[normalem]{ulem}
\usepackage{tikz}
\usetikzlibrary{positioning,angles,quotes,shapes,arrows, calc}
\usepackage{tikz-cd}
\usepackage{apacite}
\usepackage{sectsty}
\usepackage{caption, subcaption}
\usepackage[nottoc]{tocbibind}
\usepackage[colorinlistoftodos,textsize=tiny]{todonotes}
\usepackage[hidelinks,colorlinks=true,linkcolor=black,citecolor=black]{hyperref}
\newcommand{\Reals}{\mathbb{R}}
\usepackage[longnamesfirst, authoryear]{natbib}

\pgfmathdeclarefunction{gauss}{2}{%
  \pgfmathparse{1/(#2*sqrt(2*pi))*exp(-((x-#1)^2)/(2*#2^2))}%
}

\begin{document}
	\baselineskip=.22in\parindent=30pt
	
	\newtheorem{tm}{Theorem}
	\newtheorem{dfn}{Definition}
	\newtheorem{lma}{Lemma}
	\newtheorem{assu}{Assumption}
	\newtheorem{prop}{Proposition}
	\newtheorem{cro}{Corollary}
	\newtheorem{conjecture}{Conjecture}
	\newtheorem*{theorem*}{Theorem}
	\newtheorem{example}{Example}
	\newtheorem{observation}{Observation}
	\newcommand{\exm}{\begin{example}}
		\newcommand{\exmm}{\end{example}}
	\newcommand{\obs}{\begin{observation}}
		\newcommand{\obss}{\end{observation}}
	\newcommand{\cor}{\begin{cro}}
		\newcommand{\corr}{\end{cro}}
	\newcommand{\con}{\begin{conjecture}}
		\newcommand{\conn}{\end{conjecture}}
	\newtheorem{exa}{Example}
	\newcommand{\ex}{\begin{exa}}
		\newcommand{\exx}{\end{exa}}
	\newtheorem{remak}{Remark}
	\newcommand{\rmk}{\begin{remak}}
		\newcommand{\rmkk}{\end{remak}}
	\newcommand{\thm}{\begin{tm}}
		\newcommand{\nt}{\noindent}
		\newcommand{\thmm}{\end{tm}}
	\newcommand{\lm}{\begin{lma}}
		\newcommand{\lmm}{\end{lma}}
	\newcommand{\ass}{\begin{assu}}
		\newcommand{\asss}{\end{assu}}
	\newcommand{\df}{\begin{dfn}  }
		\newcommand{\dff}{\end{dfn}}
	\newcommand{\prp}{\begin{prop}}
		\newcommand{\prpp}{\end{prop}}
	\newcommand{\bqu}{\sloppy \small \begin{quote}}
		\newcommand{\equ}{\end{quote} \sloppy \large}
	\newcommand\cites[1]{\citeauthor{#1}'s\ (\citeyear{#1})}
	
	\newcommand{\eq}{\begin{equation}}
	\newcommand{\eqq}{\end{equation}}
	\newtheorem{claim}{\it Claim}
	\newcommand{\cl}{\begin{claim}}
		\newcommand{\cll}{\end{claim}}
	\newcommand{\bit}{\begin{itemize}}
		\newcommand{\eit}{\end{itemize}}
	\newcommand{\ben}{\begin{enumerate}}
		\newcommand{\een}{\end{enumerate}}
	\newcommand{\bcen}{\begin{center}}
		\newcommand{\ecen}{\end{center}}
	\newcommand{\fn}{\footnote}
	\newcommand{\ds}{\begin{description}}
		\newcommand{\dss}{\end{description}}
	\newcommand{\prf}{\begin{proof}}
		\newcommand{\prff}{\end{proof}}
	\newcommand{\cs}{\begin{cases}}
		\newcommand{\css}{\end{cases}}
	\newcommand{\ml}{\item}
	\newcommand{\lb}{\label}
	\newcommand{\ra}{\rightarrow}
	\newcommand{\tra}{\twoheadrightarrow}
	\newcommand*{\supp}{\operatornamewithlimits{sup}\limits}
	\newcommand*{\inff}{\operatornamewithlimits{inf}\limits}
	\newcommand{\nf}{\normalfont}
	\renewcommand{\Re}{\mathbb{R}}
	\newcommand*{\mmax}{\operatornamewithlimits{max}\limits}
	\newcommand*{\mmin}{\operatornamewithlimits{min}\limits}
	\newcommand*{\argmax}{\operatornamewithlimits{arg max}\limits}
	\newcommand*{\argmin}{\operatornamewithlimits{arg min}\limits}
	\newcommand{\uhr}{\!\! \upharpoonright  \!\! }
	
	\newcommand{\CR}{\mathcal R}
	\newcommand{\CC}{\mathcal C}
	\newcommand{\CT}{\mathcal T}
	\newcommand{\CS}{\mathcal S}
	\newcommand{\CM}{\mathcal M}
	\newcommand{\CL}{\mathcal L}
	\newcommand{\CP}{\mathcal P}
	\newcommand{\CN}{\mathcal N}
	\newcommand{\dee}{\mathrm{d}}
	\newtheorem{innercustomthm}{Theorem}
	\newenvironment{customthm}[1]
	{\renewcommand\theinnercustomthm{#1}\innercustomthm}
	{\endinnercustomthm}
	\newtheorem{einnercustomthm}{Extended Theorem}
	\newenvironment{ecustomthm}[1]
	{\renewcommand\theeinnercustomthm{#1}\einnercustomthm}
	{\endeinnercustomthm}
	
	\newtheorem{innercustomcor}{Corollary}
	\newenvironment{customcor}[1]
	{\renewcommand\theinnercustomcor{#1}\innercustomcor}
	{\endinnercustomcor}
	\newtheorem{einnercustomcor}{Extended Theorem}
	\newenvironment{ecustomcor}[1]
	{\renewcommand\theeinnercustomcor{#1}\einnercustomcor}
	{\endeinnercustomcor}
	\newtheorem{innercustomlm}{Lemma}
	\newenvironment{customlm}[1]
	{\renewcommand\theinnercustomlm{#1}\innercustomlm}
	{\endinnercustomlm}
	\newtheorem{innercustomdf}{Definition}
	\newenvironment{customdf}[1]
	{\renewcommand\theinnercustomdf{#1}\innercustomdf}
	{\endinnercustomdf}
	\newtheorem{innercustomex}{Example}
	\newenvironment{customex}[1]
	{\renewcommand\theinnercustomex{#1}\innercustomex}
	{\endinnercustomex}
	
	\newtheorem{innercustomprp}{Proposition}
	\newenvironment{customprp}[1]
	{\renewcommand\theinnercustomprp{#1}\innercustomprp}
	{\endinnercustomprp}
	
	\newcommand{\red}{\textcolor{red}}
	\newcommand{\blue}{\textcolor{blue}}
	\newcommand{\purple}{\textcolor{purple}}
	\newcommand{\mred}[1]{\color{red}{#1}\color{black}}
	\newcommand{\mblue}[1]{\color{blue}{#1}\color{black}}
	\newcommand{\mpurple}[1]{\color{purple}{#1}\color{black}}

	\newcommand{\AG}[1]{\textcolor{magenta}{[AG:\ #1]}}
	\newcommand{\MU}[1]{\textcolor{blue}{[MU:\ #1]}}

\newcommand*{\firstdraft}{First Version: August 15, 2023}  
\newcommand*{\seconddraft}{This Version: March  14, 2025}	
	\makeatletter
	\newcommand{\customlabel}[2]{%
		\protected@write \@auxout {}{\string \newlabel {#1}{{#2}{}}}}
	\makeatother
	
	
	\def\qed{\hfill\vrule height4pt width4pt
		depth0pt}
	\def\reff #1\par{\noindent\hangindent =\parindent
		\hangafter =1 #1\par}
	\def\title #1{\begin{center}
			{\LARGE {#1}}
	\end{center}}
	\def\author #1{\begin{center} {\large #1}
	\end{center}}
	\def\date #1{\centerline {\large #1}}
	\def\place #1{\begin{center}{\large #1}
	\end{center}}
	
	\def\date #1{\centerline {\large #1}}
	\def\place #1{\begin{center}{\large #1}\end{center}}
	\def\intr #1{\stackrel {\circ}{#1}}
	\def\R{{\rm I\kern-1.7pt R}}
	\def\N{{\rm I}\hskip-.13em{\rm N}}
	\newcommand{\cprod}{\Pi_{i=1}^\ell}
	\let\Large=\large
	\let\large=\normalsize
	
	

	\begin{titlepage}
		\def\thefootnote{\fnsymbol{footnote}}
		\vspace*{0.1in}
		\title{Robust Comparative Statics with \vspace{0.2em} \\ Misspecified  Bayesian Learning\fn{I am sincerely grateful to my committee members—Robert Anderson, Ying Chen, and M. Ali Khan—for their thoughtful comments and guidance throughout this project. I also appreciate  the comments of  three anonymous referees. I am also thankful to Abhinav Anand, Robert Barbera, Paulo Barelli, Sarvesh Bandhu, Chris Carroll, Rama Chellappa, Yujian Chen, Finn Christensen, Liuchun Deng, Haosui Duanmu, Ignacio Esponda, Marcelo Fernandez, Mira Frick, Gagan Ghosh, Steve Hamilton, Ashwin Kambhampati, Edi Karni, Jong Jae Lee, Jason Lepore, Andrew Mackenzie, Vikram Manjunath, Arina Nikandrova, Mallesh Pai, Arthur Paul Pedersen, Kevin Reffett, Metin Uyanik, Marie-Louise Vierø, Xinyang Wang, Lukasz Wozny, and Sezer Yasar for their valuable feedback and encouragement at various stages of this work. I also appreciate the feedback received from audiences at seminars and conferences, including  the 21st Annual SAET Conference, Cal Poly, IIM Bangalore, Johns Hopkins University, the XXXII European Workshop on Economic Theory at the University of Manchester, the Midwest Economic Theory Conference at the University of Rochester, and the Econometric Society Session on Misspecified Beliefs at ASSA meetings. Edited by Eduardo Azevedo.}}
		\medskip
		\author   {\Large Aniruddha Ghosh\fn{Economics Area, Orfalea College of Business, California Polytechnic State University, San Luis Obispo, CA 93401. {\bf E-mail} {aghosh10@calpoly.edu}.}}
		
	  \date{\Large \seconddraft}
			
		\vskip 0.50em
				
		\baselineskip=.15in

\smallskip
\begin{abstract} 
\nt We present novel monotone comparative statics results for steady-state behavior in a dynamic optimization environment with misspecified Bayesian learning. Building on \cite{ep21a}, we analyze a Bayesian learner whose prior is over parameterized transition models but is misspecified in the sense that the true process does not belong to this set. We characterize conditions that ensure monotonicity in the steady-state distribution over states, actions, and inferred models. Additionally, we provide a new monotonicity-based proof of steady-state existence, derive an upper bound on the cost of misspecification, and illustrate the applicability of our results to several environments of general interest.
\end{abstract}
		
		\smallskip
		
		\small{\noindent {\it Journal of Economic Literature} Classification
		Numbers: C61, D83, D90}

		\bigskip
		
		
		\bigskip

	\end{titlepage}
	
	\large

	\bigskip

	\bigskip

	\setcounter{footnote}{0}

	\large
\newpage	
\bqu
\nf \textit{In general, dynamic programs with learning are intractable, i.e., they are not solvable either analytically or numerically, when there is no separability of control and learning. The problem is not only whether a solution exists, but if a solution can be characterized and its properties studied.}\hfill{Koulovatianos, Mirman, and Santugini (2009)} 
\equ

\section{Introduction}

 Stemming from the foundational contributions of \cite{ag73} and \cite{ny91}, model misspecification in economic environments continues to be of substantive interest for economists. It acknowledges the possibility that economic agents, perhaps due to cognitive limitations or simplified perspectives, may often not know the true model of their complex economic environment and, therefore, learn about it through a set of models that are \textit{misspecified} in the sense that the true model is not contained in their set of models. A growing body of literature in economics suggests that enriching classical economic environments with misspecified economic agents aligns theoretical predictions closely with the observed behavior.\fn{For example, \cite{fa23} find that much of the anomalies concerned with forecasts of professional forecasters can be attributed to them not knowing the true model of the environment.}

Dynamic programming offers one robust framework for economists to approach discrete-time economic problems. They are commonly used to analyze instances where economic agents make decisions sequentially in an uncertain environment. The framework applies to a broad spectrum of economic models, from consumption-savings decisions for individuals to investment choices for firms. However, they are known to be intractable, no closed-form solutions, when incorporated with agents who simultaneously make decisions and learn about their environment through their models, where both the decision making and the learning activity mutually affect each other. In this paper, we provide robust predictions on the monotone comparative statics properties of an agent's \textit{steady-state} behavior within a dynamic programming environment, particularly when they are learning with potentially misspecified models.\fn{As is well-known, monotone comparative statics analysis deals with characterizing conditions for various environments, both static and dynamic, under which decision rules and solution concepts are increasing in the primitives.}

Consider the following setting, conceptualized by \cite{ep21a} (hereafter, EP), in the context of a single-agent dynamic optimization problem, a Markov Decision Process (hereafter, MDP). In each period $t=0,1,2,\ldots$, an agent observes a state $s_{t}$ and then takes an action $x_{t}$ and receives a utility, $u(s_{t}, x_{t}).$ The current action and the state together determine the evolution of the state tomorrow, $s_{t+1},$ via the true transition function, $Q (\cdot|s_{t}, x_{t}).$ The agent chooses a sequence of actions to maximize their current and the expected discounted flow of utility. The agent doesn't know the true (objective) transition function and, therefore, chooses actions based on a set of model (subjective) transition functions, $\{{Q_{\theta}, \theta\in\Theta}\},$ parameterized by $\theta,$ that is misspecified in the sense that it does not contain the true transition function. The Bayesian agent has a prior $\mu$ on the set of models $\Theta$ and, every period updates the prior using Bayes' rule $B,$ based on the current state, the chosen action, and the realized state. The agent's problem can be formulated as a  Bellman equation,
\begin{equation}
\label{bellman}
V(s, \mu)=\max_{x \in \mathbb{X}} \Bigl\{ \int_{\mathbb{S}}\{u\left(s, x, s'\right)+\beta  V\left(s^{\prime}, \mu^{\prime}\right)\}{\bar{Q}_\mu}\left(d s^{\prime} \mid s, x\right)\Bigl\},
\end{equation} where $\bar{Q}_\mu=\displaystyle\int_{\Theta} Q_\theta \mu(d \theta),\ \mu^{\prime}=B\left(s, x, s', \mu\right)$ is the next period's belief,\fn{ The Bayesian operator $B$, where for any $A \subseteq \Theta$,
$B(s, s', x, \mu)(A) = \displaystyle\int_{A} Q_{\theta}(s' | s,x) \mu(d\theta) \Big/ \displaystyle\int_{\Theta} Q_{\theta}(s' | s,x) \mu(d\theta).$} updated using Bayes' rule, and $V: \mathbb{S} \times \Delta(\Theta) \rightarrow \mathbb{R}$ is the unique solution to the Bellman equation of the agent, where $\Delta(\Theta)$ is the set of probability distributions on the parameter space, $\Theta$. 

For a misspecified agent that follows the above setting in choosing their sequence of actions and updating their sequence of beliefs over models, EP predicts its steady-state behavior in terms of a solution concept called the \textit{Berk-Nash equilibrium}. It is an equilibrium distribution over states and actions corresponding to which the misspecified agent infers the model (or a set of models) that `best-fits' the true model. In turn, given the inferred model, the chosen actions are optimal for each state. Furthermore, the steady-state is stationary in the sense that equilibrium distribution over states and actions leads to a stationary Markov process over the states. The `best-fit’ formalization is in the sense of a minimum weighted Kullback-Leibler divergence, with weights determined by the equilibrium distribution over states and actions.\fn{The Kullback-Leibler (KL) divergence measures the difference between two probability distributions in terms of relative entropy; in our case, between the model-implied process and the true process; see \cite{co05} for details.} Given this prediction of the long-run behavior of a misspecified Bayesian agent in terms of a Berk-Nash equilibrium and its corresponding inferred model, we ask the following comparative statics question: how do the equilibrium objects respond to changes in the economic primitives? Phrased differently, what are the requirements on the primitives of the economic environment so that a misspecified agent’s stationary distribution and corresponding best-fit models exhibit monotone comparative statics properties?

This question is of general theoretical interest for several reasons. First, models of dynamic programming incorporated with learning, whether correctly specified or misspecified, are known to be intractable.\fn{With  learning incorporated in dynamic programming models, the agent is an active learner and processes incoming data in order to infer the models of their unknown environment. At the same time, they also take decisions. These two functions of the agent are intertwined.} In an important paper on learning in stochastic growth models, \cite{kms09} (hereafter, KMS) note that the intractability arises primarily for two reasons: (a) the curse of dimensionality problem, which occurs because of incorporation of beliefs about the models in the state space, thus making the state space very large and computationally cumbersome, and (b)  with Bayesian updating, the prior and the posterior over models potentially belong to different families, thereby hampering analytical and computational tractability.\fn{As KMS and EP note, the curse of dimensionality does not arise if the beliefs are over a finite parameter set. However, for parameter spaces such as the real line, it is a cause for concern. Also note that, unlike KMS, in our framework, choosing actions also affects the flow of information, thus  further complicating matters.} By focusing on the long-run prediction for a misspecified dynamic program with a precise characterization of the distribution over states and actions, as well as the limiting posterior (beliefs) over models, we abstract away from the intractability problems and thereby make predictions about their steady-state properties even when their closed-form solutions may not exist. Thus, in this paper, we expand the theory of dynamic programming with misspecified learning, potentially rendering it applicable across a wide spectrum of economic domains. 

Second, comparative statics for misspecified learning environments are typically not well-understood in the existing literature on misspecification in economic theory. This traces itself to the issue that with concurrent decision making and learning, there is often no separability between the two, as is illustrated in Equation (1). We tackle this problem by focusing on the monotone comparative statics properties and recognizing that steady states for such models are fixed points of a well-defined equilibrium mapping. This insight allows us to appeal to a relatively newer literature of powerful order-theoretic techniques, tailored for dealing with the monotonicity of fixed points for non-lattice spaces (\cite{aj15}). A technical challenge that lies with misspecified MDPs is that the underlying spaces containing such fixed points are typically not lattices in any natural order, and therefore, a wide array of popular lattice-dependent techniques (\cite{hp92, to98}) are not applicable.  Finally, our results for comparative statics do not necessitate any specific knowledge of the functional forms of any environment primitives, and therefore are \textit{robust}, and hold for a very general class of dynamic programming environments with misspecified models.\fn{The reader may be interested in the following discussion: https://stats.stackexchange.com/questions/274815/why-should-i-be-bayesian-when-my-model-is-wrong. \cite{up03} discuss the contrast between Savagean and Knightian approaches to tackle model misspecification concerns. The interested reader is referred to their p. 2469 for further reading.} 

To interpret our framing, we discuss examples that cover three important economic environments.  The first example concerns Bayesian inference with misspecified AR(1) models. The agent's set of models is misspecified in the sense that although the true process is AR(1), the innovations of the process are non-normally distributed and, therefore, are not contained in their set of AR(1) models with Gaussian innovations. We characterize the Berk-Nash equilibrium for this example, and show that despite being misspecified, the Bayesian agent correctly infers the persistence parameter of the true process at the steady state. The next two examples are based on EP, albeit with slight modifications for interpretation and analytical tractability.  In the second example, we study a canonical problem of effort provision in a dynamic effort-task problem where the agent seeks to learn their ability while exercising control over the stochastic outcome of a task by choosing their effort levels.  The agent's models are misspecified since they incorrectly postulate that tomorrow's outcome depends only on their ability and not on today's outcome. In contrary, the true dynamics are determined by both ability and current outcome. The Berk-Nash equilibrium for this case is the steady-state frequency of the outcome, either a success or a failure, along with the agent's inference of their ability at the steady state. We characterize the dependence of these equilibrium objects on the economic primitives of the environment, such as the cost of exerting higher effort or the objective probability of success, and study their associated comparative statics.  The final example considers an agent that optimally chooses its consumption and savings in the presence of preference and productivity shocks. The agent seeks to learn about the return of their wealth process but does so through a set of misspecified models that do not consider the correlation between the two shocks. Here, the Berk-Nash equilibrium is the long-run perceived distribution of the wealth process along with their corresponding inferred parameter of the return on that process. However, this case has no analytical solution; therefore, making our results most valuable for such settings.\fn{Additionally, in Remark 2 we show the applicability of our results to a dynamic version of a setting involving an over-confident agent who has a misspecified model of their ability, based on \cite{hks18}.}

We now turn to the specificity of our results. Theorem 1 establishes the existence of a Berk-Nash equilibrium for a misspecified MDP with compact Euclidean state and action spaces. EP and \cite{adgk22} (hereafter, ADGK) provide proofs for the existence of a Berk-Nash equilibrium with finite and infinite state and actions, respectively. ADGK applies recent advances in non-standard analysis to present existence theorems for misspecified environments with infinite action and state spaces (compact-metric) and unbounded payoff functions. In contrast, here we focus on compact Euclidean spaces, where existence also emerges as a by-product of the monotonicity assumptions. The novelty of our existence result lies in exploiting the non-lattice structure of our economic environment. In general, distributions over states and actions are not lattices under most natural orders; therefore, to show the existence of the Berk-Nash equilibrium which in turn is a distribution, we rely on a novel application of techniques tailored for non-lattice spaces. These techniques, originating in \cite{sm71}, have been pioneered in \cite{aj15} for establishing the existence and comparative statics results in infinite-horizon large dynamic economies.

After furnishing the existence result for the Berk-Nash equilibrium, we utilize these techniques for Theorems 2-4 and give sufficient conditions for which the Berk-Nash equilibrium and the corresponding inferred best-fit model respond monotonically to the primitives of the environment. We formalize a notion of a positive shock for a misspecified dynamic optimization problem, defined as shocks to the primitives that increase strategies for any given beliefs over models, and under mild assumptions establish in Theorem 2 that the least and the greatest inferred best-fit model at the steady state, responds monotonically to a change in the economic primitives.  Theorem 2 further provides a prediction for the monotonicity of the Berk-Nash equilibrium in the usual stochastic order, when a misspecified agent is hit with positive shocks to their primitives. In Theorem 3, we show that the least and the greatest inferred best-fit model responds monotonically to an \textit{increase}  in the set of models, where an increase is defined in the strong-set order in the space of parameterized models.  Theorem 4 drops the usual stochastic order on the underlying space and extends Theorem 2 to spaces with increasing and convex orders.\fn{As \cite{ckk21} note in the context of individual choices, the strong set order proves to be an appropriate notion; see the references therein for further details.} 

Our final result is of a slightly different flavor from the previous results. In Theorem 5, we perform a welfare comparison of steady-state learning in a correctly specified vis-\`a-vis a misspecified learning environment.  While welfare under misspecified learning is always weakly lower than that under correct learning, we provide an upper bound on the cost of misspecification, formalized as the difference between the two instances and outline its dependence in terms of the primitives of the environment.\fn{We also remark on the potential usefulness of this bound for computational applications concerning Berk-Nash equilibria.}

\medskip

\nt\textbf{Related literature.} This paper belongs to the growing literature on learning with misspecified models and its implications on the properties of the learning behavior.

Within the dynamic programming framework, following EP and ADGK, this paper works with a general framework (Euclidean) to model MDPs with misspecified learning for infinite environments.\fn{See \cite{pu94} for illustrations covering operations research, engineering, and many other allied fields. Also, see \cite{ru94} for an extensive survey on MDPs and associated structural estimation methods. The results in this paper could potentially be admissible in an extended learning version of the partially observable MDP framework of \cite{sa18}; see the references therein for an overview of that literature.} MDPs are commonly used to analyze instances involving agents making decisions sequentially in an uncertain environment and, therefore, have found many applications within natural and social sciences. Within the economics literature, one of its earliest uses is traceable in \cite{ahm51}, where an inventory holder maximizes some given objective (profits, revenue, net utility) by choosing an inventory policy subject to stochastic product demand and other random fluctuations to the primitives of the problem.\fn{This work has important precursors in \cite{abg49}.} Since then, MDPs have spanned both micro and macro environments, covering many important economic applications; these include (i) investment with adjustment costs under uncertain demand (\cite{lp71}), (ii) one-sector optimal economic growth with uncertainty (\cite{bm72}), (iii) equilibrium search and unemployment (\cite{lp74}), and (iv) asset prices in a pure exchange economy (\cite{lu78}).\fn{Other prominent examples include models of portfolio choice under uncertainty (\cite{ph62, ls69}) and  business cycle models (\cite{kp82, lp83}).} In a recent paper,  \cite{sa18} presents a framework for an MDP in which agents possess a `cloud' of models and allows for considering model ambiguity and the potential misspecification within an MDP. 

This paper also contributes to the understanding of the comparative static properties of the limit posterior of a Bayesian inference process, specifically a Markov process. In his foundational paper on inference with misspecified models, devoid of any actions, \cite{be66} demonstrates that for a misspecified Bayesian agent learning about a parameter through a series of independently and identically distributed (i.i.d.) signals, the posterior concentrates asymptotically on those set of models (referred therein as the asymptotic carrier) where the Kullback-Leibler divergence is minimal with respect to the true model.\fn{See also \cite{bm98} and \cite{sh09} for extensions to non-iid environments. The reader should note that Berk’s asymptotic carrier is independent of the prior distribution.} In this paper, we address a misspecified MDP, a non-i.i.d. environment, that combines decision-making and inference but heavily relies on Berk's characterization to abstract away from the dynamics of the updating process. This approach enables us to focus on the comparative statics of the steady-state behavior for both the Berk-Nash equilibrium and the corresponding asymptotic carrier, and contribute to the inference literature by outlining the dependence of the inferred models on the primitives of the decision making and the learning environment.\fn{This also has comparative statics implications for a series of papers that follow the papers of \cite{hu67} and \cite{wh82}. These papers use maximum likelihood techniques for inference and show that an agent’s inference converges to models minimizing the Kullback–Leibler divergence from the true distribution, consistent with the limits proposed in \cite{be66}.}

The topic of misspecification in economic environments has been a subject of active research area.\fn{Among many other prominent works, some notable ones include \cite{ki75}, \cite{je05}, and \cite{hs11}. The reader is referred to the references therein for a broader view of the literature.} In an influential paper, \cite{ep16ecma} provides a framework for a static game-theoretic setting that relaxes the assumption that agents have a correct view of the game's environment. This  formulation is further extended to the dynamic programming environment in EP for a finite (states and actions) setting and by ADGK for an infinite setting.  While there has been influential work on social learning problems as in  \cite{fii20} and \cite{bh21}, properties of asymptotic learning in a misspecified MDP environment are still not well-studied. The results in this paper contribute to the understanding of such properties, in terms of the comparative statics behavior of misspecified agents.

Naturally, this work is informed by the substantive literature on monotone comparative statics. The theory of monotone comparative statics deals with characterizing conditions under which the optimizing  behavior of agents leads to solution concepts (and invariant distributions) being monotonic in the primitives of the environment.\fn{Some important references include \cite{ms94},  \cite{am96}, \cite{to98}, \cite{hu03}, \cite{drw18}, and  \cite{li21}. \cite{am18} is a useful guide for some of the history and also of the more recent advances in the comparative statics literature.} In an influential paper, \cite{hp92} use the tools of monotone comparative statics to analyze stationary dynamic optimization problems with lattice environments.  Their results turn out to be of limited use for obtaining our theorems. Our proofs instead exploit the non-lattice structure of our setting
and, therefore, depend on the techniques developed in \cite{sm71} and \cite{aj15}. We adapt and apply these techniques in a novel way to misspecified MDPs with Bayesian learning.

Lastly, we calculate the cost of misspecification in terms of the discrepancy between the expected discounted welfare under correctly specified and misspecified Bayesian learning for an MDP and provide an upper bound on the discrepancy between the two quantities. We then characterize its comparative statics properties with respect to the primitives. This is partly inspired by \cite{sa00}'s work on testing the accuracy of numerical solutions in dynamic models. We argue that the one-way bound on welfare that we provide  is useful for computational purposes of such equilibria.

\medskip
\nt\textbf{Outline.} The paper is organized as follows. Section 2 sets up the  framework for a misspecified MDP and outlines the necessary prerequisites and techniques for the comparative statics analysis. Section 3 offers three examples to illustrate our setting. The main results of this paper are presented in Section 4, while Section 5 completes the formal analysis of the examples. Section 6 compares welfare between correctly specified and misspecified  settings.   Section 7 concludes by discussing how these results can be extended and related to different  settings. Appendix \ref{mathpreliminaries} to this paper contains the necessary mathematical preliminaries, and the proofs of the results are presented in Appendix B.

\section{General Framework}

We begin by framing the environment for a Markov Decision Process (MDP) with model misspecification. While the conceptual framework is patterned after the finite (states and actions) environment of EP, our setting is based on Euclidean spaces since it allows infinite settings that naturally feature many important economic applications. After outlining the contents of a misspecified MDP and the relevant Berk-Nash equilibrium concept, we provide an overview of the order-theoretic methods that are instrumental for our results. A refresher for these concepts and methods is provided in Appendix \ref{mathpreliminaries} for the reader's convenience.

 At the start of each period $t=0,1,2, \ldots,$ the agent observes a state realization, $s_{t}\in \mathbb S,$ and then chooses an action, $x_{t}\in \mathbb  X.$ Given a transition probability function $Q(\cdot|s_{t},x_{t}),$  the state and action together determine the distribution of the next period state, $s_{t+1}.$ The per-period payoff function is a mapping $u: \mathbb S\times \mathbb S \times \mathbb X \rightarrow \mathbb R.$ The agent maximizes expected discounted utility (discount factor $0<\beta<1$) by choosing a feasible sequence of policy functions $\{x_{t}\}_{t=1}^{\infty}$ that solves the following problem,
\eq
\label{sequence}
V(s_{0})=\max_{\{x_{t}\}_{t=0}^{\infty}} \mathbb{E}_{Q}\Bigg[\sum_{t=0}^{\infty} \beta^{t} u(s_{t}, x_{t}, s_{t+1})\Bigg], \ t=0,1,2, \ldots.
\eqq

\nt Following (\ref{sequence}), the Bellman equation for the agents' sequential
decision problem is formulated as,
\begin{equation}
\label{bellman}
V(s)=\max_{x \in \mathbb{X}} \Bigl\{\int_{\mathbb{S}} \{u\left(s, x, s'\right) +\beta V\left(s^{\prime}\right)\} Q\left(d s^{\prime} \mid s, x\right)\Bigl\},
\end{equation}
where $V,$ the value function, is the unique solution to (\ref{bellman}). Corresponding to this $V,$ the optimal policy correspondence $G$ is given by,  
\eq
\label{policy}
G(s)\equiv \argmax_{x \in \mathbb{X}}\displaystyle \Bigl\{\int_{\mathbb{S}} \{u\left(s, x, s'\right)+\beta V(s')\} Q(\dee s'|s, x)\Bigl\}.
\eqq

\nt We summarize this environment in the following definition of a MDP.\fn{The framework can be extended with minor modifications to allow for the dependence of feasible set of actions on the state variable; see the \cite{ep15wp} working paper  for details.}

\df\label{defMDP} \nf
A Markov Decision Process is a tuple $\langle \mathbb S,\mathbb X,q_0,Q, u\rangle$, where 
\nf(i) the state space $\mathbb S\subseteq \mathbb R^{m}$ is a compact metric space with Borel $\sigma$-algebra $\mathcal B(\mathbb S),$ (ii) the action space $\mathbb X\subseteq \mathbb R^{n}$ is a compact metric space with Borel $\sigma$-algebra $\mathcal B(\mathbb X),$ (iii) the initial distribution of states $q_0$ is a probability measure on $\mathbb S$, (iv) $Q: \mathbb S\times \mathbb X\to \mathcal M_{1}(\mathbb S)$ is a transition probability function, where $\mathcal M_{1}(\mathbb S)$ denote the set of probability measures on $\mathbb S$, and (v) $u: \mathbb S\times \mathbb S\times \mathbb X \to \Reals$ is the per-period payoff function.\fn{$\mathcal{M}_{1}(\mathbb S)$ is the space of finite probability measures on $(S, \mathbb{B}(S))$ endowed with the weak-$*$ topology.} 
\dff

\nt We next define a Subjective Markov Decision Process. It adds to the tuple, a set of subjective transition functions, $\{Q_{\theta}\}_{\theta\in \Theta},$ parameterized with $\theta.$ We refer to the parameter space $\Theta$ as the set of \textit{models}.

\df \nf
\label{smdp}
A Subjective Markov Decision Process (hereafter, SMDP) is an MDP, $\left\langle\mathbb{S}, \mathbb{X}, q_{0}, Q\right.$, $u \rangle$, and a non-empty family of transition probability functions, $\mathcal{Q}_{\Theta}=\left\{Q_{\theta}: \theta \in \Theta\right\}$, where each transition probability function $Q_{\theta}: \mathbb{S} \times \mathbb{X} \rightarrow \mathcal M_{1}(\mathbb S)$ is indexed by a parameter value $\theta \in \Theta\subseteq \mathbb R$.\fn{Our focus here is on an uni-dimensional parameter space $\mathbb{R}$ to align with applications commonly found in the literature, however, the results in this paper are applicable to $\mathbb{R}^d,$ with slight modifications. The reader is also referred to the discussion on multi-dimensional parameter spaces in Section 7.} A SMDP is said to be \textit{misspecified} if $Q\notin \mathcal Q_{\Theta}.$
\dff

\nt Notice that under this definition, an SMDP  could be misspecified in several ways. For instance, the true transition function and the set of model transition functions may pertain to dissimilar families of probability distributions. Alternatively, even if the true model and set of models belong to the same family of distributions, they can be misspecified if the support, the range of possible values of the parameter, of the models is different from that of the true distribution.\fn{See Examples 1 and 2 for an instance for the former, and Example 3 for the latter. \cite{hd19} explore more on forms of model misspecification and their relevance to econometric methods with macroeconomic models.} Our next definition requires the primitives of the SMDP to satisfy certain \textit{regularity} conditions.

\df\label{regsmdp}\nf
A \textit{regular} SMDP satisfies the following conditions.
\begin{enumerate}[{\normalfont (i)}, topsep=1pt]
 \setlength{\itemsep}{-2pt}
    \item (Continuity) The mappings $(s, x) \to Q(\cdot \mid s, x)$ and $(\theta, s, x) \to Q_{\theta}(\cdot \mid s, x)$ are continuous in the Prokhorov metric, and the density function $D_\theta\left(s^{\prime} \mid s, x\right)$ is jointly continuous on the set 
\[
\left\{\left(\theta, s^{\prime}, s, x\right) : Q(s, x) \text{ is dominated by } Q_\theta(s, x)\right\},
\]
where $\big(D_{\theta}(s' \mid s, x)\big)$ is the Radon-Nikodym derivative of $Q$ with respect to $Q_{\theta},\ \theta \in \Theta.$\fn{The Prokhorov metric is used to measure the closeness of probability measures based on their tightness and weak convergence. For two transition functions \( Q_1 \) and \( Q_2 \), the Prokhorov metric is defined as:
\[
d(Q_1, Q_2) = \inf \{ \epsilon > 0 : Q_1(A) \leq Q_2(A^\epsilon) + \epsilon \text{ and } Q_2(A) \leq Q_1(A^\epsilon) + \epsilon, \ \forall A \in \mathcal{B}(\mathbb{X}) \},
\]
where \( A^\epsilon \) is the \( \epsilon \)-neighborhood of \( A \), given by: $
A^\epsilon = \{ x \in \mathbb{X} : d(x, a) < \epsilon \text{ for some } a \in A \}.$ This metric is particularly well-suited for compact state spaces.}

        \item (Absolute continuity) There is a dense set $\hat{\Theta}\subset \Theta$ such that $Q(\cdot| s, x)$ is absolutely continuous with respect to $Q_{\theta}(\cdot| s, x)$ for all $\theta\in \hat{\Theta}$ and $(s, x)\in \mathbb S\times \mathbb X$.
    \item\label{KLint} (Uniform integrability) For every compact set $S'\subset S$, there exists some $r>0$ such that $\big(D_{\theta}(\cdot|s, x)\big)^{1+r}$ is uniformly integrable with respect to $Q_{\theta}(\cdot| s, x)$ over the set $\hat\Theta$.\fn{The density function $D_{\theta}(\cdot|s, x)$ is the Radon-Nikodym derivative of $Q$ with respect to a model, $Q_{\theta}.$ The uniform integrability is satisfied if the density functions $D_{\theta}(\cdot|s, x)$ are uniformly bounded over the set $\{(\theta, s, x): Q(\cdot| s, x)$ \mbox{is dominated by} $Q_{\theta}(\cdot| s, x)$\}. For example, given any two Gaussian distributions with distinct variances, the Radon-Nikodym derivative of the one with the larger variance with respect to the other is unbounded. }
      \item (Compactness) The parameter space $\Theta$ is a compact metric space.
\end{enumerate}
\dff

\nt

Condition (i) is a standard technical condition and requires the transition functions, both true and subjective, to be continuous. Condition (ii) requires that there always exists some model $Q_{\theta}$ that accounts for every observation from the true distribution, $Q.$ Condition (iii) places an uniform integrability requirement to deal with distributions with infinite support. Lastly, condition (iv) requires the parameter space $\Theta$ to be compact. This is an essential requirement for the existence of best-fit models as will be evident in the next definition. We next measure the extent of misspecification in terms of the well-known measure of relative entropy, the Kullback-Leibler divergence.

\df\nf For a given true transition function $Q$ and model transition function $Q_{\theta},$ the Kullback-Leibler divergence, $\mathcal{D}_{\mathrm{KL}},$ of $Q_{\theta}$ with respect to $Q$ is defined as, 
\eq
\label{defKLD}
\mathcal{D}_{\mathrm{KL}}\big(Q(s, x), Q_{\theta}(s, x)\big)=\mathbb{E}_{Q(\cdot|s, x)}\left[\ln \big(D_{\theta}(s'|s,x)\big)\right].
\eqq  Then for any distribution over states and actions, $m\in \mathcal M_{1}(\mathbb S\times \mathbb X),$ and  model parameter, $\theta\in \Theta$, the \textit{weighted} Kullback-Leibler divergence is a mapping $K_{Q}: \mathcal M_{1}(\mathbb{S}\times \mathbb{X})\times \Theta\to \bar{\mathbb R}_{\geq 0}$ such that 
\eq
K_{Q}(m,\theta)=\int_{\mathbb{S}\times \mathbb{X}}\mathcal{D}_{\mathrm{KL}}\big(Q(s, x), Q_{\theta}(s, x)\big)\\m(\dee s, \dee x).\fn{$\bar{\Reals}$  denotes the extended real line, equipped with the one-point compactification topology.}
\eqq
\nt The set of closest parameter values given a distribution $m\in \mathcal M_{1}(\mathbb S\times \mathbb X)$ and for a given true transition function $Q$ is the set, $ \Theta(m; Q)=\argmin_{\theta\in \Theta}K_{Q}(m,\theta).$\fn {$\mathcal{M}_{1}(\mathbb S\times \mathbb X)$ is the space of finite probability measures on $(\mathbb S\times \mathbb X, \mathcal B(\mathbb S\times \mathbb X))$ endowed with the weak-$*$ topology. The density function $D_{\theta}(s'|s, x)$ is jointly continuous on the set $\{(\theta, s', s, x)\!: Q(s, x)\text{is dominated by}\ Q_{\theta}(s, x)\}$. We follow the standard convention in that $\ln(0)\cdot 0=0$ and integral of infinity over a set of measure $0$ is $0$.}
\dff

\nt That is, given a distribution $m$ over states and actions and a true transition function $Q,$ the \textit{best-fit} set $\Theta(m; Q)$ is the set of those parameter values  that minimize the weighted relative entropy between true transition function and the parameterized  model transition  functions in $\Theta.$ Our focus in this paper is on the Berk-Nash equilibrium. We now define it. 

\df
\nf A probability distribution $m^{*}\in\mathcal M_{1}(\mathbb S\times \mathbb X)$ is a Berk-Nash equilibrium of the regular-SMDP  if there exists a belief $\mu^{*}\in \mathcal M_{1}(\Theta)$ such that the following conditions hold.
\begin{enumerate}[{\normalfont (a)}, topsep=1pt]
 \setlength{\itemsep}{-2pt}
    \item  For all  states and actions, $(s,x),$ that are in the support of $m^{*}$, action $x$ is \textit{optimal} given state $s$ in the MDP($\bar{Q}_{\mu^{*}}$), where $\bar{Q}_{\mu^{*}}=\displaystyle\int_{\Theta}Q_{\theta}\mu^{*}(\dee \theta)$.
     \item  For a given true transition function $Q,$ beliefs $\mu^{*}$ are \textit{restricted} over the best-fit set of parameters, that is, $\mu^{*}\in \mathcal M_{1}(\Theta(m^{*}; Q))$.
    \item  For all $A\in \mathcal B(\mathbb S)$, $m^{*}_{\mathbb S}(A)=\displaystyle\int_{\mathbb S\times \mathbb X}Q(A|s, x)m^{*}(\dee s, \dee x)$, where $m^{*}_{\mathbb S}$ denote the \textit{invariant} marginal measure of $m^{*}$ on $\mathbb S$. 
\end{enumerate}
\label{berknash}
\dff

\nt  The Berk-Nash equilibrium 
is a probability distribution $m^{*}$ over the states and actions, supported by equilibrium beliefs $\mu^{*}$ over the best-fit set of parameterized models. The beliefs $\mu^{*}$ are optimal given the data generated by the steady-state distribution over states and actions, $m^{*}.$ In turn, the distribution $m^{*}$ over states and actions is such that actions, conditioned on the state, are subjectively optimal given the equilibrium beliefs $\mu^{*}$ over the best-fit models.  For instance, in the case of a correctly specified MDP, where the true transition function $Q$ is part of the set of models $\mathcal{Q}_{\Theta}$, the concept of Berk-Nash equilibrium implies that in the steady state for a Bayesian learner, beliefs concentrate on the true transition function, and the actions are optimal given this belief. The equilibrium concept effectively reduces to the well-known stationary solution for a MDP$(Q)$.

However, if the SMDP is misspecified, that is, if $Q$ is not contained within the set $\mathcal{Q}_{\Theta}$, then in Definition (\ref{berknash}), condition (a) asserts that action $x$ is  optimal in the MDP$(\bar Q_{\mu^{*}})$ when the equilibrium transition function is a weighted combination of model transition functions, with weights given by $\mu^{*}$. Condition (b) mandates that these beliefs $\mu^{*}$ are determined by minimizing the weighted relative entropy between the true transition function and the model transition functions, where the weights are the Berk-Nash equilibrium, $m^{*}$. Lastly, condition (c) stipulates that the marginal distributions over states, $m^{*}_{\mathbb S}$ is an invariant distribution. Notice that in condition (c), the invariant distribution is dependent on the optimal actions $x$ via the true transition function, $Q$.  EP show that under moderate conditions, the learning illustration with misspecified models sketched in Equation (1) converges to the Berk-Nash equilibrium.\fn{EP, in their paper, demonstrate the usefulness of their equilibrium concept by providing a learning foundation for finite spaces (Theorems 2 and 3), under either positive visitation or identification. The identification condition also holds in our setting, courtesy of Assumption 3. \cite{adgkv2} have partially addressed the convergence problem for infinite spaces in an ArXiv version (v2) of their manuscript, available here: [\href{https://arxiv.org/pdf/2206.08437v2}{https://arxiv.org/pdf/2206.08437v2}; see Section C.2 in Appendix, pages 48–56]. For infinite (but compact) state and action spaces, they demonstrate that convergence holds under identification, but only in the total-variation norm on the space of state and action distributions.}  The equilibrium serves as a prediction of the steady-state behavior in  where $m^{*}$ is the steady-state distribution over states and actions, and $\mu^{*}$ is the limit of the sequence of posteriors of a Bayesian learner with misspecified models.

Our focus is on the comparative statics of the Berk-Nash equilibrium and the associated best-fit set with respect to the primitives $P$ of the environment, where $P=<u, \beta, Q, Q_{\Theta}, \Theta>$ are our objects in the collection of primitives. Given any primitive $p\in P,$ we next adjust our value function, the optimal policy correspondence, and the best-fit set  to allow for their dependence on the primitive.  At the steady state solution, agent has a belief $\mu^{*}$ over their set of best-fit models and solves a stationary dynamic programming problem where the value function $V:\mathbb S\times \mathcal M_{1}(\Theta) \times P \rightarrow \mathbb R$ is determined by the following functional equation, where $\bar Q_{\mu^{*}}=\displaystyle\int_{\Theta} Q_{\theta}\mu^{*}(d\theta),$

\begin{equation}
\label{stationarybellman}
V(s,\mu^{*}, p)=\max_{x\in \mathbb X}  \Bigl\{\int_{\mathbb{S}} \{u\left(s, x, s'\right) 
+\beta V(s', \mu^{*}, p)\}\bar Q_{\mu^{*}}(ds'|s,x),\Bigl\}\mbox{ and }
\end{equation}

\nt corresponding to $V$ and $\mu^{*},$ the stationary optimal policy correspondence $G$ is given by,  
\eq
\label{stationarypolicy}
G(s,\mu^{*}, p)\equiv \argmax_{x\in \mathbb  X} \Bigl\{\int_{\mathbb{S}} \{u\left(s, x, s'\right)+\beta  V(s', \mu^{*}, p)\}\bar Q_{\mu^{*}}(ds'|s,x)\Bigl\}.
\eqq

\noindent Therefore, in line with Definition \ref{berknash}, the set of Berk-Nash equilibriua for a given primitive $p$ is given by the set of fixed points of the equilibrium mapping $T$, \begin{equation} \Lambda(p) \equiv\{ (m^{*},\mu^{*}) \in \mathcal M_{1}(\mathbb S\times \mathbb X)\times \mathcal M_{1}(\Theta) : z \in T(z, p)\},  \end{equation}
\nt where $T: Z\times P\tra 2^{Z}$ is a set-valued function on the space of probability measures on states and actions and the set of parameters, $Z=\mathcal M_{1}(\mathbb S\times \mathbb X)\times \mathcal M_{1}(\Theta)$ and $P=<u, \beta, Q, Q_{\Theta}, \Theta>$ are our primitives.

For inquiring  the comparative statics behavior of the equilibrium objects in $\Lambda(p)$ with respect to the primitives of the environment, we need to define corresponding orders for the equilibrium objects and the primitives. For vectors $x$ and $y$ in $\mathbb R^{n}$,  we use the following convention: ``$x\geq y$" means $x_i\geq y_i$  in every component,  ``$x> y$" means $x\geq y$ and  $x\neq y$, and ``$x\gg y$" means $x_i>y_i$  in every component. A real-valued function,  $f: X \rightarrow \mathbb{R},$ is said to be increasing if for $x\geq y$ \mbox{in the component-wise order}, $f(x)\geq f(y),$ and convex if its domain is a convex set and for all $x, y$ in its domain, and all $\lambda \in[0,1]$, we have
$f(\lambda x+(1-\lambda) y) \leq \lambda f(x)+(1-\lambda) f(y).$\fn{The interested reader is referred to \cite{ss07} for more details concerning univariate and multivariate orders.} We  follow \cite{ss07} in ranking probability measures on the state and action space, and the parameter space. 

\df[\cite{ss07}] 
Let \( \mathcal{M}(X) \) denote the space of probability measures on a compact subset \( X \subseteq \mathbb{R}^k \) for some finite \( k \in \mathbb{N} \). For any two measures, $\mu$ and $\nu$ in $\mathcal M(X),$ 
 \smallskip
 \ben[{\nf (i)}, topsep=1pt]
\setlength{\itemsep}{-2pt}
\item  $\mu$ usual order stochastically dominates $\nu$, $\mu\succsim_{st} \nu,$ if $\displaystyle\int f(x) \mu(\dee x) \geq \displaystyle\int f(x) \nu(\dee x)$, for any measurable, bounded, and increasing real-valued function $f.$ 
\item $\mu$ increasing and convex-order stochastically dominates $\nu$, $\mu\succsim_{icx} \nu,$ if $\displaystyle\int f(x) \mu(\dee x) \geq \displaystyle\int f(x) \nu(\dee x)$, for any measurable, bounded, increasing, and convex real-valued function $f.$
\een
\dff

\smallskip

We rely on well-known orders for the primitives. For instance, the discount factor $0<\beta<1$ is ranked in the natural order, as in higher patience implies a higher $\beta.$ The true transition function $Q$ can be ordered in multiple ways, including the usual stochastic order, the convex order, or the increasing convex order.\fn{Convex order in this case, would refer to changes in true distribution in the sense of mean-preserving spreads. For two distributions, $Q_{1}$ and $Q_{2},$ distribution $Q_{2}\succsim_{cx} Q_{1}$ iff $\displaystyle\int f(s) Q_{2}(\cdot|s, x) \geq \displaystyle\int f(s) Q_{1}(\cdot |s, x)$ for every convex function, $f,$ and for all $(s,x)\in \mathbb S\times \mathbb X.$} The parameter set $\Theta$ is ranked in the strong-set order, where $\Theta_{2}$ is  greater than or equal to $\Theta_{1}$ if, for any $\theta$ in $\Theta_{2}$ and any $\theta'$ in $\Theta_{1}$, the maximum of $\{\theta, \theta'\}$ belongs to $\Theta_{2}$ and the minimum of $\{\theta, \theta'\}$ belongs to $\Theta_{1}$. Changes in the primitives of the utility function refer to changes that impact utility levels such as variations in risk aversion or other factors, depending on the specific applications.

After defining all the necessary prerequisites for analyzing the comparative statics of the fixed points in Equation (9), the next logical step is to apply the standard lattice theoretic methods as used in (\cite{hp92}, \cite{to98}) and derive the monotonicity of the fixed points, the Berk-Nash equilibrium and the corresponding inferred model, in terms of the primitives. However, there is a technical challenge here. Notice that for our equilibrium mapping $T,$ the underlying space on which it is defined is not a lattice. That is, the space of probability measures over states and actions (and parameters) are not lattices in any natural order.\fn{Let $\mathcal M(X)$ denote the space of probability measures defined on a compact subset of $X \displaystyle\subset \mathbb{R}^n$. Although even if $X$ is a lattice, the poset $(X,\succsim_{st})$ is not a lattice as pointed in \cite{kko77}. For e.g., for $(\mathbb R^{2},\succsim_{st})$ let $p_{1}= 0.5 (\epsilon_{a}+\epsilon_{b}), p_{2}= 0.5 (\epsilon_{a}+\epsilon_{c}),p_{3}= 0.5 (\epsilon_{c}+\epsilon_{b}),p_{4}= 0.5 (\epsilon_{a}+\epsilon_{d}),$ where $a=(0,0), b=(0,1), c=(1,0), d=(1,1).$ Then, both $p_{3}$ and $p_{4}$ are supremum, which is a contradiction.} Thus, standard lattice theoretic methods are of no use for our misspecified MDP setting. Therefore, to establish the monotonicity of the fixed points, our results rely on a rather novel application of a technical framework that originated in \cite{sm71} and was pioneered in the context of large dynamic economies by \cite{aj15}, for non-lattice spaces. 

\rmk 
\normalfont{Our setting differs from that of \cite{aj15} in several key aspects. Their work examines the steady-state outcomes of an infinite-horizon dynamic economy with a continuum of agents experiencing idiosyncratic (exogenous) shocks, whereas we focus on the steady-state outcomes of a single agent facing an infinite-horizon dynamic problem, where shocks depend on the agent's action choices and are therefore endogenous. Furthermore, our setting features a Bayesian agent learning with misspecified models of the shock process, unlike \cite{aj15}, which assumes that agents know the true distribution of idiosyncratic shocks, thereby removing the need for any learning component. As a result, their techniques require adaptations to suit our framework (see Auxiliary Lemmas 1 and 2).}
\rmkk

Towards this end, we shall show (Auxiliary Lemmas 1 and 2 in Appendix B) that our equilibrium mapping $T$ is monotonic in the sense defined below.

\df[\citet{sm71}]\nf  Let $X$ and $Y$ be sets  equipped with some partial order $\succsim$, and $P,$ a partially ordered set. A correspondence $T: X \times P \rightarrow 2^Y$ is Type I monotone in $x$ for each $p$ if for all $x_1 \succsim x_2$ and $y_2 \in T\left(x_2,p\right)$, there exists $y_1 \in$ $T\left(x_1,p\right)$ such that $y_1 \succsim y_2,$ and  Type II monotone  if for all $x_1 \succsim x_2$ and $y_1 \in T\left(x_1,p\right)$, there exists $y_2 \in$ $T\left(x_2,p\right)$ such that $y_1 \succsim y_2$. A correspondence $T: X \times P \rightarrow 2^Y$ is Type I monotone in $p$ for each $x$ if for all $p_1 \succsim p_2$ and $y_2 \in T\left(p_2,x\right)$, there exists $y_1 \in$ $T\left(p_1,x\right)$ such that $y_1 \succsim y_2,$ and  Type II monotone  if for all $p_1 \succsim p_2$ and $y_1 \in T\left(p_1,x\right)$, there exists $y_2 \in$ $T\left(p_2,x\right)$ such that $y_1 \succsim y_2$. $T$ is Type I (Type II) monotone if it is Type I (Type II) monotone in both $X$ and $P.$\fn{Smithson uses the term `multifunction' in his paper. All the proofs in this section are given for Type I monotonicity. The proofs for Type II monotonicity follow analogously. It is important to point out here that while the statement for Theorem 1 is correct in \cite{sm71}, its proof is wrong; the correct proofs are in \cite{ho86}.}
\dff

By placing mild monotonicity structure on our environment, we shall show that the Berk-Nash equilibrium map $T$ is Type I and Type II monotone. And then by appealing to an existence result in \cite{sm71} for non-lattice spaces, we shall show that the set of fixed points is non-empty.\fn{The reader may wish to pause and review the material in the mathematical preliminaries  in Appendix A.}  Finally, the following  result in \cite{aj15} will play a critical role in transfer the monotonicity of the equilibrium mapping $T$ onto the set of its fixed points. We explain in more detail the applicability of these results in Section 4. 

\begin{theorem*}[Acemoglu-Jensen (2015)]
\medskip
\label{thm-aj15}
Let $X$ be a compact topological space equipped with a closed partial order $\succsim,$ $P$ a partially ordered set of primitives, and $T: X \times P \rightarrow 2^{X}$ be upper hemicontinuous for each $p \in P$. Define the fixed-point correspondence,
$$\Lambda(p)=\{x \in X : x \in T(x, p)\}.$$ Then if $T$ is Type I (Type II) monotone in $p$, so is $\Lambda$ in $p.$
\end{theorem*}

\section{Motivating Examples}

We now present three examples of misspecified environments that illustrate the framework of this paper. Our first example is that of a misspecified AR(1) inference process where actions have no role. It is purely an inference problem.\fn{Per se, there is no decision process involved here. One could think of an economic situation where the flow of utility is constant over time, and therefore, actions play no role in either inference, nor, payoffs.} Examples 2 and 3 are more substantive in that the choice of actions plays a role in learning, and vice-versa. They are based on EP and cover effort provision problems with misspecified effort-task dynamics and consumption-savings problem with misspecified wealth process, respectively. While Example 2 has an explicit analytical solution and is, therefore, used to outline the mechanics for comparative statics properties of
a Berk-Nash equilibrium, the solution for Example 3 is intractable and, therefore, the most substantive for the applicability of this paper’s techniques.\fn{We omit the full analysis of Examples 2 and 3 since they are already covered in the aforementioned papers (EP and ADGK).}  

\ex[Inference of an AR(1) process]\nf 
In this example, we characterize the Berk-Nash equilibrium of an inference process when the set of AR(1) models, parameterized by $\theta,$ is misspecified.  Let the state space $\mathbb S=\mathbb R.$ Suppose the state variable $s_{t+1}$ evolves via the true transition function, $Q(\cdot|s_{t}).$ Let the true process $Q(\cdot|s_{t})$  be an AR(1) process with parameter $0<|\rho|<1$ and with the innovations distributed as  a two-component mixture normal distribution with one component $(\mu_{1},\sigma_{1}^{2})$ and the other component $(\mu_{2},\sigma_{2}^{2}),$ and  given by, 
\eq
s_{t+1}=\rho s_{t}+\xi_{t+1}, \hspace{0.5cm} \xi_{t+1} \sim 0.5 F_{(\mu_{1},\sigma^{2})} + 0.5 F_{(\mu_{2},\sigma^{2})}
\eqq 
\nt where $F$ denotes the cumulative density function for a normal distribution. The components have different means ($\mu_{1}\neq \mu_{2})$ but identical variances  $(\sigma_{1}^{2}=\sigma_{2}^{2}).$ An agent is equipped with a compact set $\Theta$ of parameterized AR(1) models, $Q_{\theta} (\cdot|s_{t})$, with Gaussian noise, and the process,
\eq
s_{t+1}=\theta s_{t}+\xi_{t+1}, \hspace{0.5cm} \xi_{t+1} \sim N(0,\sigma^{2}).
\eqq

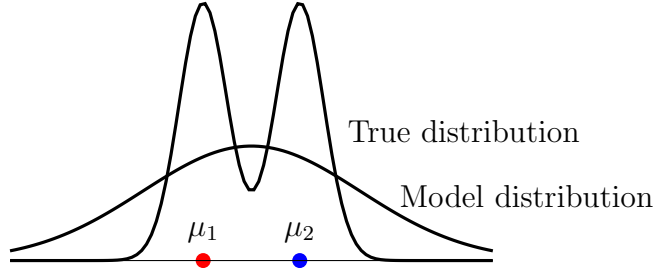
\begin{figure}[h]
\label{lapfigure}
\centering
\begin{tikzpicture}
\begin{axis}[
  no markers, domain=0:10, samples=100,
  axis lines*=left, xlabel=, ylabel=$y$,
  every axis y label/.style={at=(current axis.above origin),anchor=south},
  every axis x label/.style={at=(current axis.right of origin),anchor=west},
  height=5cm, width=8cm,
  xtick=\empty, ytick=\empty,
  enlargelimits=false, clip=false, axis on top,
  grid = major,
  axis y line=none 
  ]
  \addplot [very thick,black] {0.5*gauss(4,0.5)+0.5*gauss(6,0.5)};
 \node[black] at (axis cs: 9.4, 0.2) {True distribution}; 
  \node[black] at (axis cs: 10.7, 0.1) {Model distribution}; 
    \addplot [very thick,black] {gauss(5,2.25)};
  \node[circle, fill=red, inner sep=2pt, label=above:$\mu_{1}$] at (axis cs: 4, 0) {};
  \node[circle, fill=blue, inner sep=2pt, label=above:$\mu_{2}$] at (axis cs: 6, 0) {};
\end{axis}
\end{tikzpicture}
\caption{While the innovation process in the true distribution is  mixture-normally distributed  with distinct means and identical variances, resulting in a bi-modal distribution, the innovation processes in the agent's models conform to a normal distribution.}
\end{figure}

The agent's models are misspecified since the true distribution is not normally distributed, and therefore, is not in the set of models, $Q_{\Theta}$. Following Definition (\ref{berknash}), the Berk-Nash equilibrium is the stationary distribution $m^{*}_{S}$ over the set of states (no actions), implied by the true  process, $Q(\cdot|s_{t}),$ with the corresponding best-fit AR(1) parameter, $\theta^{*}=\displaystyle\int_{\mathbb S} \hat\theta(s)m^{*}(ds),$ where, 
$$\hat\theta(s)\equiv \argmin_{\Theta} \mbox{KL}(Q|Q_{\theta})=\argmin_{\Theta}\mathbb E_{Q}\bigg[\dfrac{\ln Q(\cdot|s)}{\ln Q_{\theta}(\cdot|s)}\bigg].$$

\nt Now, given that the AR(1) models are Gaussian and therefore log-concave, the state-dependent best-fit model $\hat\theta(s)$ is  uniquely determined.\fn{It is worth mentioning here  that Berk's characterization of the limit posterior is for an i.i.d. process. Here conditioned on the state $s,$ the AR(1) process behaves as i.i.d., and therefore, $\hat\theta(s)$ could be interpreted as the posterior to which the agent converges if they were to receive infinite state-realizations, all i.i.d., given the state is fixed.} Since the innovations in the models are Gaussian, $\hat\theta(s)$ is the least-squares minimizer with respect to the true transition $Q(\cdot|s_{t}),$ and solves the following, 
\eq\label{statewise}
\hat\theta(s)E_{Q(\cdot|s)}(s^{2})=E_{Q(\cdot|s)}(s's).
\eqq

The best-fit inferred AR(1) parameter has the following form, 
$$\theta^{*}=\displaystyle\int_{\mathbb S}\hat\theta(s)m^{*}_{\mathbb S}=\rho+\displaystyle\int_{\mathbb S}\dfrac{(\mu_{1}+\mu_{2})}{2s}m^{*}_{\mathbb S}.$$ 

Now, suppose $\mu_{1}+\mu_{2}=0.$ That is,  the mean of the components of the true distribution cancel each other. In this case, a Bayesian agent, starting from any prior on their models $Q_{\Theta}$, would eventually converge in the Berk-Nash equilibrium to the true persistence parameter $\rho$ and yet continue to remain misspecified. For this case, the comparative statics of the steady state inferred  $\theta^{*}$ with respect to the AR(1) parameter of the true process is one-to-one in a trivial way.\fn{For the case when $\mu_{1}+\mu_{2}\neq 0,$  the expression for $\theta^{*}$ is potentially intractable since the expectation with respect to the Berk-Nash equilibrium $m^{*}$ may simply not exist. The inspiration for the mixture normal distribution comes from a paper by \cite{cs23} who look at learning when the set of likelihoods is misspecified.} 
\exx

\ex[Effort provision with unknown ability (EP)]\label{dem}\nf In this setting of a dynamic effort provision problem, we follow EP in characterizing the Berk-Nash equilibrium. The equilibrium for this instance is a probabilistic prediction for the agent's long run outcome to be either a success or a failure, along with their inference of their ability. This example is related to the growing literature on effort provision, as modeled in \cite{dw22}.

Each period, an agent chooses either to put high ($H$) or low ($L$) effort, $x_t \in \mathbb{X}=\{L, H\}, L<H$ in a task.  The task then has two outcomes, either it fails (0) or succeeds (1), that is, $s_{t+1} \in \mathbb{S}=\{0,1\}.$ The payoffs are as follows,
 \begin{equation*}
    u(x_{t}, s_{t+1})   = \begin{cases}
 s_{t+1,} & \text{$x_{t}=L,$} \\
s_{t+1} - c, & \text{$x_{t}=H.$}
\end{cases} \end{equation*}
\nt  In case the effort is low, the agent's payoff is the state realization, while if the effort is high, in addition to the state realization, they also incur a cost $c$ for their high effort. Under the true process $Q(\cdot|s_{t}, x_{t})$, if the agent puts in the high effort,  the probability of the task being a success is 1, irrespective of the outcome of the task in the current state. That is, $Q(1 \mid s_{t}, H)= 1.$  With low effort, however, the probability of success tomorrow is dependent on the outcome today; $Q(1 \mid 0, L)= q_{0},$ if it is a failure,  and  $Q(1 \mid 1, L) = q_1$, if it is a success.  Following EP, we assume that   $0<q_{0}< 1-c <q_{1}<1.$ From a routine calculation in EP, the optimal correctly specified policy function is the following,
\begin{equation*}
x(s_{t})   = \begin{cases}
 H, & \text{$s_{t}=0,$} \\
L, & \text{$s_{t}=1.$}
\end{cases} 
\end{equation*}
\nt That is, a failure today, pushes the agent to work harder (since $q_{0}<1-c$), while a success makes them put lower effort (since $1-c<q_{1}$). 

However, the agent has a set of misspecified models $\mathcal Q_{\Theta}=\{Q_{\theta}\}$, parameterized by $\theta\in\Theta=[0,1],$ where $Q_{\theta}(1|s,H)=1$ and $Q_{\theta}(1|s,L)=\theta$ for all $s\in \{0,1\}.$ The agent's models correctly captures the true dynamics,  if they were to put in higher effort.  However, under lower effort, their models imply that the chances of success are  independent of the state today, and is denoted by the parameter $\theta.$ Following a growing literature on effort provision that models an agent's control over their future success, based on their effort and ability, we novelly interpret the parameter $\theta$ as an index for the agent's ability level. That is, the agent is someone who thinks that when the effort is high, then their ability plays no role in determining the outcome tomorrow since higher effort guarantees success. Whereas, when the effort is low, it is only the ability that matters for the successful outcome, with higher ability implying higher chances of success. 

For the Berk-Nash equilibrium, following EP, we shall focus on the unique mixed-strategy equilibrium wherein the choice of actions are independent of the current state, as implied by the  misspecified models. A routine calculation shows that under certain parameterizations, a unique Berk-Nash equilibrium exists wherein the agent plays a mixed strategy, conditioned on the state, $m^{*}(L|0)=m^{*}(L|1)= \dfrac{q_{1}-(1-c)}{c(q_{1}-q_{0})},$  and correspondingly infers their ability, $\theta^{*}=1-c.$\fn{The steady-state marginal distribution is given by  $m^{*}_{S}(1)=\dfrac{1-c-q_{0}}{q_{1}-q_{0}},$ and $m^{*}_{S}(0)=1-m_{S}(1).$ Further, $\theta_Q(m)=\left(1-m_{\mathbb{S}}(1)\right) q_0+m_{\mathbb{S}}(1) q_1$. The reader is referred to the EP paper (pp. 726-728) for the full numerical details. } 

\exx

\ex[Savings with misperceived wealth process (EP, ADGK)]\label{ocs} \nf In this example, we demonstrate that there are environments wherein there is no clear analytical solution for the Berk-Nash equilibrium and its corresponding inferred model, a feature most common to dynamic programming problems.

In each period, an agent realizes wealth $y_{t},$ an i.i.d. preference shock $z_{t},$ and subsequently chooses to save $x_t\in [0, y_t]= \mathbb X\subseteq\mathbb{R}_{+}.$ The utility in period $t$ is given by $u(y_t, z_t, x_t) = z_t \ln (y_t - x_t)$. The state variables, denoted as $s = (y, z)$, belong to $\mathbb{S} = \mathbb{R}_{+} \times [0, 1]$. The evolution of wealth $y_{t+1}$ in the following period is governed by the following process,
\[\ln y_{t+1} = \alpha^* + \beta^* \ln x_t + \varepsilon_t,\]
\nt where the unobserved productivity shock, $\varepsilon_t=\gamma^* z_t + \xi_t$, with $\xi_t$ following a standard normal distribution, $\xi_t \sim \mathcal{N}(0,1),$ and $z_{t}$ following a uniform distribution, with mean $E(z).$ Here, $\beta^*$ denotes the true return on one additional unit of log of saving in terms of the log of wealth. We assume that it lies between 0 and 1, so that the true Markov process is stationary. Following EP, we shall assume that the preference and the productivity shocks are positively correlated, $\gamma^* > 0$. For an agent that knows all the primitives of the environment, solving for the Bellman as in Equation (\ref{bellman}), the correctly specified optimal policy function is given by, $x^{*} = A_{z}(\beta^{*})y,$ where $A_{z}(\beta^{*}) = \displaystyle\dfrac{\delta\beta^{*}E(z)}{(1-\delta\beta^{*})z + \delta\beta^{*} E(z)},$ where $0<\delta<1$ is the discount factor.

\medskip

However, the agent's set of models postulates the following wealth process, $\ln y_{t+1}=\alpha+\beta \ln x_t+\varepsilon_t,$
where $\varepsilon_t \sim N(0,1)$, thus ignoring the correlation between the productivity and the preference shocks. So while the true process is $Q\left(y^{\prime}, z^{\prime} \mid y, z, x\right)$ is such that $y^{\prime}$ and $z^{\prime}$ are independent, $y^{\prime}$ has a log-normal distribution with mean $\alpha^*+\beta^* \ln x+\gamma^* z$ and unit variance, and $z^{\prime} \sim U[0,1],$ the misspecified process is $Q_\theta\left(y^{\prime}, z^{\prime} \mid y, z, x\right)$ is such that $y^{\prime}$ and $z^{\prime}$ are independent, $y^{\prime}$ has a log-normal distribution with mean $\alpha+\beta \ln x$ and unit variance, and $z'\sim U[0,1].$

For this case, the Berk-Nash equilibrium is characterized by an optimal policy function, $x^{m}=A_z\left(\beta^m\right) y=\displaystyle\dfrac{0.5 \delta \beta^m}{\left(1-\delta \beta^m\right) z+0.5 \delta \beta^m} y$, where there exists a corresponding inferred model $\beta^m \in\left(0, \beta^*\right)$. That is, the misspecified agent underestimates the return on her saving, $\beta^{m}<\beta^{*},$ and therefore, undersaves in the Berk-Nash equilibrium, $x^{m}<x^{*}.$\fn{The intuition behind this is that a higher preference shock $z$ is associated with a lower saving proportion, $A_{z} (\beta^{m}),$ since it leads to a lower inferred belief of return on saving, $\beta^{m},$ since the preference and productivity shocks are positively correlated.} Notice however that there is no closed-form expression for $\beta^{m}$  that outlines its dependence on the environment primitives, and therefore, how the inferred return, the equilibrium policy function, and the invariant distribution respond to the primitives of the environment is an open question.\fn{From EP: The inferred model at the stationary Berk-Nash equilibrium  solves the following equation, 
\begin{equation}
\hat \beta (A_{z}(\beta))=\beta^*+\gamma^* \frac{\operatorname{Cov}\left(z, \ln A_{z}(\beta)\right)}{\operatorname{Var}\left(\ln A_z(\beta)\right)+\operatorname{Var}(\ln Y)},
\end{equation}
where the covariance and variance are taken with respect to the true distribution, $Q(y', z'|y, z).$ Under the parametric assumptions of $\gamma^{*}>0,$ and $Cov (z, A_{z})<0$, EP establish that there exists a $\beta^{m},$ the inferred model in the Berk-Nash equilibrium. Further, in EP, notice that the inferred model depends on the proportion, $A_{z},$ instead of the distribution over states and actions, which is unlike the case  required as per the conditions of Berk-Nash equilibria.}

\exx

\section{Main Results}

We now turn to our main results that are on the existence and the comparative statics of Berk-Nash equilibrium for a misspecified dynamic optimization problem. We invoke techniques from the fixed point literature (\cite{sm71,aj15}) and, in doing so, impose further structure on our regular SMDP. While assumptions 1 and 2 are useful for establishing increasing policy selections, assumptions 3 and 4 give the required monotonicity and identification properties for our models. Throughout, we shall assume that $Q$ is monotone. The proofs for all the results are in Appendix B.

\ass
\label{assumption1}
The state and action spaces are lattices, and for any primitive $p\in P,$  $u(s, s', x, p)$ is supermodular in $(s, x)$ and increasing in current state $s$ and future state $s'.$
\asss

Our first assumption is rather standard and emphasises a lattice structure on the state, action, and parameter spaces. As evinced by the examples in this paper and in general, for many economic applications,  the state, action, and parameter spaces are a subset of the real line $\mathbb R,$ for which the lattice assumption is trivially satisfied in the natural order. Further, we assume that the payoff function is supermodular in current state and action and is increasing in the future state variable. A payoff 
 function $u: \mathbb S\times \mathbb X \rightarrow \mathbb{R}$ is supermodular if, for all $(s_{1},x_{1}), (s_{2},x_{2}) \in \mathbb S\times \mathbb X, u((s_{1},x_{1}) \vee (s_{2},x_{2}))+u((s_{1},x_{1}) \wedge (s_{2},x_{2})) \geq u(s_{1},x_{1})+u(s_{2},x_{2}).$ This implies that states and actions exhibit complementarity in the sense that marginal contribution to the payoff of increasing the action increases with a higher state. The supermodularity in states and actions and monotononicity (in state) of the payoff function  alongwith our next assumption guarantees that optimal policy correspondence is increasing in the state variable, for any given model distribution $\mu,$ and primitive $p\in P.$

\ass
\label{assumption2}
The following holds true for all models in the family of models, $\mathcal{Q}_{\Theta}=\left\{Q_{\theta}: \theta \in \Theta\right\}.$ For any increasing real-valued function $f(\cdot),$

\ben[{\nf (i)}, topsep=1pt]
\setlength{\itemsep}{-2pt}
\item $Q_{\theta}$ is stochastically increasing in $(s,x)$ i.e.$\displaystyle\int_{\mathbb S} f(s') Q_{\theta}(ds'|s,x)$ is increasing in $(s,x).$ 
\item $Q_{\theta}$ is stochastically supermodular in $(s,x)$ i.e. $\displaystyle\int_{\mathbb S} f(s') Q_{\theta}(ds'|s,x)$ is supermodular in $(s,x).$ \een
\asss

Notice that this assumption is exclusively on the set of models.  It requires the model transition functions to be stochastically increasing and stochastically supermodular in states and actions. This implies that for every model $Q_{\theta},$ a higher current state and action increases the probability of observing a higher state in the next period and that an incremental amount of action increases this probability, the higher the state is in the current period.\fn{ For example, consider the following AR(1) process, $s_{t+1}=\theta s_{t}+\epsilon_{t+1},$ where $\epsilon_{t+1}$ is distributed normally with mean 0 and variance $\sigma^{2}.$ Then for every $\theta,$ $Q_{\theta}$ is stochastically increasing. See Examples 2 and 3 for illustrations on stochastic supermodularity.} Assumptions 1 and 2 are consistent with the structure one requires for a  correctly specified environment and the reader is referred to Theorem 3.9.2 in \cite{to98} for a similar set of assumptions. For a given model distribution $\mu,$ and primitive $p,$ the above two assumptions guarantee an increasing optimal policy correspondence $G(s,\mu, p).$ 

We next assume that given any observable endogenous data, i.e. given any distribution $m$ over states and actions, generated by agent's endogenous learning and decision making process as in Equation (1), the  best-fit parameter is point-identified. That is, given $m,$ the best-fit set is singleton and therefore, the parameter is uniquely determined. 

\ass
\label{assumption3} For any given $m \in \mathcal{M}_{1}(\mathbb{S} \times \mathbb{X})$, we assume that the SMDP $(Q, Q_{\Theta})$ is point-identified, meaning that $\theta, \theta' \in \Theta(m; Q)$ implies $\theta = \theta'$. Moreover, we assume this holds for all $m \in \mathcal{M}_{1}(\mathbb{S} \times \mathbb{X})$.\fn{The reader is referred here to \cite{le19} for an exposition on identification problem in economics. }
\asss

The identification assumption is satisfied for many applications in the existing misspecification literature and also plays an important role in the convergence of the steady state behavior of an agent in a SMDP to the Berk-Nash equilibrium; see examples in \cite{ep16ecma}, EP, and, \cite{ep21b}. For our comparative statics purposes, identification ensures that we always get a unique selection $\theta_{Q}(m)$ from the best-fit set $\Theta(m; Q).$  A function $\theta_{Q}(m),$  maps a distribution $m$ into a real number in $\Theta$ is said to be increasing in the usual stochastic order $\succsim_{\mathrm{st}}$,  if $\theta_{Q}(m_{2}) \geq \theta_{Q}(m_{1})$ whenever $m_{2} \succsim_{\text {st }} m_{1}.$ For our last assumption, we require the weighted Kullback-Leibler divergence to exhibit the single-crossing property in the parameter $\theta$ and distribution $m$.

\ass $K_{Q}(\theta;m)$ satisfies the single crossing property in $(\theta;m),$ that is, for $\theta_{1}\leq \theta_{2},$ and for $m_{2}\succsim_{st} m_{1},$ $\delta(m_{1})=K_{Q}(\theta_{2},m_{1})-K_{Q}(\theta_{1},m_{1})\geq (<) 0$ implies $\delta(m_{2})=K_{Q}(\theta_{2},m_{2})-K_{Q}(\theta_{1},m_{2})\geq (<)  0.$\fn{It is important to note that as long as the divergence ``measure" satisfies a form of single crossing, as outlined in Assumptions 4 and 5, the comparative statics results remain valid. In this regard, the KL divergence serves more as an example than a critical feature for monotonicity. However, our emphasis on the KL divergence stems from the fact that the Berk-Nash solution concept identifies it as a potential limit of the misspecified learning process. Alternatively, if another divergence measure were to characterize the limit, it would need to satisfy the single crossing property for the comparative statics results to hold. I thank a referee for this observation.    }
\label{assumption4}
\asss

\nt The assumption above is necessary and sufficient to imply that the best-fit set  $\Theta(m; Q)$ (coupled with Assumption 3, the inference function $\theta_{Q}(m)$), is increasing in the strong-set order on the real line.\fn{This is similar to assuming \textit{monotone} Bayesian updating of the kind in \cite{to05} where stochastically higher states lead to higher posterior probabilities of the parameter. Also, since the parameter spaces are in $\mathbb R,$ quasi-supermodularity is trivially satisfied; see \cite{ms94}.} However, as is well-known in the comparative statics literature, it can be difficult to verify for single-crossing differences.  Therefore, a sufficient and easy to check condition to guarantee an increasing best-fit function or set  is to require that the models follow a \textit{expected} log-likelihood property.

\smallskip
\nt{\textbf{Sufficient Condition 1}:} For any two models $\theta_{1},\theta_{2}\in \Theta,$ such that $\theta_{1}<\theta_{2},$ define the log-likelihood ratio, $L(s'|s,x)=\ln(D_{\theta_{2}}(s'|s,x))-\ln(D_{\theta_{1}}(s'|s,x)),$ where $D_{\theta}(s'|s,x)$ are the Radon-Nikodym derivatives, introduced for measuring KL divergences in (\ref{defKLD}). Then the models are said to follow the expected likelihood ratio property  if the expectation of $L$ with respect to the true distribution $Q(\cdot|s,x)$ is increasing in both state and action, $(s,x).$\fn{\cite{ro87} contains a useful discussion regarding the existence of the expectation of the log-likelihood ratio statistic. For instance, if the true, and the model distributions belong to the Gaussian family, this statistic would always exist.}

We can now state our first result on the existence of a Berk-Nash equilibrium for a regular SMDP with infinite states and actions.

\thm
\label{thmexistence}
Under assumptions 1-3, every regular SMDP $(Q,\mathcal Q_{\Theta})$ with a bounded and continuous utility function has a Berk-Nash equilibrium and the set of such equilibria is compact.
\thmm

EP and ADGK provide proofs for the existence of a Berk-Nash equilibrium in an SMDP with finite and infinite environments, respectively. While the former proves the theorem for finite states and actions,  ADGK takes the finite result of EP as given and extends it to more naturally appealing instances of infinite environments (compact-metric spaces), using novel tools in non-standard analysis. In contrast, our proof of existence result relies on the assumed monotonicity and identification properties of the structure of our problem, given we are housed in the Euclidean space. It is important to mention here that we do rely on the regular SMDP definition in ADGK, most importantly on part (iv) of Definition \ref{regsmdp}, because it allows us to consider distributions that have unbounded Radon-Nikodym derivatives. This is particularly relevant when dealing with distributions over infinite spaces, such as the AR(1) process.\fn{While our examples focus on unbounded state spaces, the results in this paper apply to compact state and action spaces, which do not directly impact the comparative statics properties.} For unbounded state spaces, one continues to rely on ADGK's structure and existence results. Further, the set of Berk-Nash equilibria will be compact and in particular, there will always exist the least and the greatest Berk-Nash equilibrium, $m^{*}$ and corresponding beliefs $\mu^{*}$ over the best-fit inferred models.\fn{The argument for the least and greatest follows from Theorem 4 in \cite{aj15} for the non-lattice case. Also, see Footnote 9 (pp. 1389) in \cite{hp92} for several instances of compact subsets of measures in economic problems.}  

Following \cite{aj15}, we next define a positive shock for a SMDP. Notice that this positive shock is defined for any $p\in P=<u, \delta, Q, Q_{\Theta}, \Theta>$ and therefore, appeals to any primitive $p$ that can be considered.

\df
\normalfont For any given belief $\mu,$ a change in a primitive of the SMDP from $p_{1}$ to $p_{2}$ is a \textit{positive (negative)} shock if $G(s,\mu, p_{2})$ is greater than $G(s, \mu, p_{1})$ in the strong set order. That is,
for all $y_{1} \in G\left( s, \mu, p_{1}\right)$ and $y_{2} \in G\left(s, \mu, p_{2}\right),$ the join $y_{1} \vee y_{2} \in G\left(s, \mu, p_{2}\right)$ and the meet $y_{1} \wedge y_{2} \in G\left(s, \mu, p_{1}\right).$ Further, for a given primitive $p\in P$, a change in beliefs over models in the \textit{usual order stochastic dominance sense }from $\mu_{1}$ to $\mu_{2}$ is a \textit{positive (negative)} shock if $G(s,\mu,p)$ is ascending in $\mu$ from $\mu_{1}$ to $\mu_{2},$ that is, if  $y_{1}\vee y_{2}\in G(s,\mu_{2},p)$ and $y_{1}\wedge y_{2}\in G(s,\mu_{1},p)$ for all $y_{1}\in G(s,\mu_{1},p)$ and $y_{2}\in G(s,\mu_{2},p).$
\dff

We now state the main result of this paper.

\thm
\label{thmpositive}
Suppose assumptions 1-4 hold. Then a positive shock to the primitives of the regular SMDP will lead to an
increase in the least and the greatest equilibrium best-fit models.\fn{It is important to note here that we are assuming that the constraint on the parameter set is non-binding.}  Further, a positive shock to the primitives will lead to 
\begin{enumerate}[{\normalfont (a)}, topsep=1pt]
 \setlength{\itemsep}{-2pt}
\item an increase in the  least and greatest Berk-Nash equilibrium in the usual stochastic order dominance if changes in beliefs over models are positive shocks, and
\item a decrease in the least and greatest Berk-Nash equilibrium in the  usual stochastic  order  dominance if changes in beliefs over models are negative shocks.
\end{enumerate}
\thmm

\nt Theorem 2 highlights the two-way interaction between decision making and learning for monotone comparative statics behavior in misspecified dynamic optimization problems with learning. Under assumptions 1-4 on the regular SMDP, it  predicts that a positive shock to any of its primitives will lead to an increase in the least and greatest equilibrium beliefs over the best-fit parameterized models. Furthermore, a positive (negative) shock will lead to an increase in the least and the greatest Berk-Nash equilibrium, the steady state distribution over states and actions, if changes in beliefs over models is a positive (negative) shock. For instance, if there is a unique Berk-Nash equilibrium, as in the examples here and in the literature, with a corresponding inferred model, then Theorem 2 gives a clear prediction for the comparative statics of the equilibrium objects.

\rmk
\normalfont
\cite{hks18}, henceforth HKS, identify an interesting result of self-defeating learning in misspecified environments. They examine how a myopic agent, with overconfident beliefs—unrealistically high expectations—about their ability, engages in a learning process that leads them away from optimal actions. Technically, the HKS framework is not directly applicable here, as it is not a Markov Decision Process (MDP) per se. However, we adapt its basic structure to formulate it as an MDP, demonstrating how the key assumptions in HKS align with ours, and deriving the comparative statics results accordingly.\fn{I am deeply grateful to an anonymous referee for urging me to make this connection.}

In each period \( t \in \{1, 2, 3, \ldots\} \), the agent produces an observable output according to the law of motion:
\[
q_{t+1} = Q\bigl(x_t, a, \Phi\bigr) + \varepsilon_{t+1},
\]

\noindent where \( x_t \in [\underline{x}, \bar{x}] \) is the action chosen by the agent at time \( t \), \( a \in \mathbb{R} \) represents the agent’s constant ability, and \( \Phi \in [\underline{\phi}, \bar{\phi}] \) is a fixed but unobservable fundamental parameter. The term \( \varepsilon_{t+1} \) represents random noise, which is assumed to be independent and identically distributed (i.i.d.), following the assumptions specified in HKS. The agent is assumed to be overoptimistic about his ability: although his true ability is \( A \), he believes his ability is \( \tilde{a} \), where \( \tilde{a} > A \). We define the degree of overconfidence as \( \Delta = |\tilde{a} - A| \). Given this, the agent updates his beliefs about the fundamental parameter \( \Phi \) in a Bayesian manner. He discounts future outcomes with a factor \( \delta \in (0, 1) \), and in each period, chooses a current action \( x_t \) along with a strategy for future actions, contingent on the history of the game and his learning of the fundamental, in order to maximize his discounted expected output. The agent’s objective is thus to solve the following dynamic optimization problem:
\[
\max_{x_t, x_{t+1}, \dots} \mathbb{E}_t \left[ \sum_{t=0}^{\infty} \delta^t \, q_{t+1} \right],
\]
\noindent where the expectation is taken with respect to the agent’s beliefs about the future evolution of the state \( \Phi \) and the corresponding output \( q_{t+1} \), and subject to the law of motion given by the equation above.

The steady state of this dynamic optimization problem is characterized by a Berk-Nash equilibrium, which consists of a steady-state distribution over the state variable and a corresponding action choice \( x^* \), together with an equilibrium belief \( \Phi^* \) about the fundamental. In equilibrium, the agent's Bellman equation satisfies
\begin{equation} \label{eq:value-function}
V\bigl(q_t, a, \Phi^*\bigr) = \max_{x^{*} \in [\underline{x},\bar{x}]} \; \mathbb{E}\Bigl[\, q_{t+1} + \delta \, V\bigl(q_{t+1}, a, \Phi^*\bigr) \Bigr],
\end{equation}
where the expectation is taken over the distribution of \( \varepsilon_{t+1} \). Furthermore, the equilibrium belief \( \Phi^* \) must be consistent with the agent's subjective model. In particular, \( \Phi^* \) satisfies the condition
\begin{equation} \label{eq:KL-minimization}
Q\bigl(x^*, A, \Phi\bigr) = Q\bigl(x^*, \tilde{a}, \Phi^*\bigr),
\end{equation}
which arises as the solution to the problem of minimizing the Kullback-Leibler (KL) divergence between the true and the subjective distributions.\fn{The reader is referred to Lemma 7, page 1190 in HKS.}

We prove that in the non-myopic misspecified MDP version of HKS, higher overconfidence results in lower equilibrium effort. This follows from the assumptions in HKS that correspond to our Assumptions 1-4. First, HKS’s Assumption 1, $Q_{ax} \leq 0$, translates in our framework to a change in the primitive (ability) that qualifies as a negative shock due to submodularity. Second, the optimal effort level is always monotonic (increasing) in the change in fundamental, which translates to a change in beliefs about the fundamental in a usual-order stochastic sense, thus a positive shock. Third, for any effort level, there is a unique fundamental consistent with the output produced. This follows from $Q_{\phi} > 0$ in HKS, satisfying Assumption 3 in our paper. Finally, since $Q_{x\phi} > 0$ in HKS, this implies that optimal action and the fundamental are monotonic in each other.\fn{We assume, without loss of generality, that the noise $\epsilon_{t+1}$ follows a normal distribution and, therefore, satisfies regularity. Readers are referred to page 1163 in HKS for a listing of these assumptions.}
\qed
\rmkk

While the full proof of the main result is provided in the Appendix, we provide a brief proof sketch below that outlines the main points of our argument.

\medskip

\nt\textbf{Short proof sketch:}
The proof of Theorem \ref{thmpositive} follows a three-step proof structure as in \cite{aj15}.  The first step involves showing that under Assumptions 1 and 2, and for any fixed model distribution $\mu$, the set of Berk-Nash equilibriua $m$ obtained via the fixed points of the equilibrium mapping $T$ will be Type I and Type II increasing in the primitives $p.$  Using the set of fixed points in the first step, the second step involves constructing a mapping $\hat\theta$ that for any model distribution $\mu$ and primitive $p,$ gives a set of model distributions.  This mapping is constructed from part (b) of Definition \ref{berknash} wherein for each of the Berk-Nash equilibrium $m$ obtained in step 1, the construction gives a set of $\mu's.$ It is the fixed points of this map that are our equilibrium distribution given $p$. The \textit{third} and \textit{final} step shows that if the least and the greatest selections of the mapping $\hat\theta$ are increasing in primitives, then the associated fixed points are increasing in $p$.

\rmk 
\nf
\cite{aj15} gives several sufficient conditions to identify positive shocks for their environment. All of their results (Lemmas 1-3) translate to our case,\fn{The reader is referred to pp. 604-605 in \cite{aj15}.} albeit with some moderation for the endogenous dynamic programs. For example, a change in the primitive $p$ that influences the decision problem through the utility function, such as the risk-aversion parameter, is a positive shock if the utility function $u(s,x,p)$ has increasing differences in $x$ and $p.$ Similarly, under Assumptions 1 and 2, an increase in discount factor $\beta$ is a positive shock.
\rmkk

Our next theorem is on the inference of a Bayesian learner in a misspecified MDP who undergoes an \textit{expansion} of their set of models in the strong-set order.

\thm
Suppose the hypothesis in Theorem 1 continue to hold. If a change in beliefs over models in the usual stochastic order is a positive shock, then an increase in the parameter set under the strong set order leads to an increase in the least and the greatest equilibrium best-fit models. 
\thmm

Our last assumption modifies the single crossing property for instances when distributions over states and actions are ordered in the increasing and convex stochastic order. Our final theorem addresses the monotonicity of Berk-Nash equilibrium for the increasing and convex order.

\ass $K_{Q}(\theta;m)$ satisfies the single crossing property in $(\theta;m),$ for increasing convex order $\succsim_{icx}$ if, for $\theta_{1}\leq \theta_{2},$ and for $m_{2}\succeq_{icx} m_{1},$ $\delta(m_{1})=K_{Q}(\theta_{2},m_{1})-K_{Q}(\theta_{1},m_{1})\geq (<) 0$ implies $\delta(m_{2})=K_{Q}(\theta_{2},m_{2})-K_{Q}(\theta_{1},m_{2})\geq (<)  0.$
\label{assumption5}
\asss

\nt{\textbf{Sufficient Condition 2}:} For any two models $\theta_{1},\theta_{2}\in \Theta,$ such that $\theta_{1}<\theta_{2},$ define the log-likelihood ratio, $L(s'|s,x)=\ln(D_{\theta_{2}}(s'|s,x))-\ln(D_{\theta_{1}}(s'|s,x)).$ Then the models are said to follow the expected likelihood ratio property in increasing and convex order if the expectation of $L$ with respect to the true distribution $Q$ is increasing and convex in state and actions, $(s,x).$

\thm 
Suppose assumptions 1-3 and assumption 5 holds. Then a positive (negative) shock to the primitives will lead to an increase in the least and greatest Berk-Nash equilibrium in the increasing convex order if change in beliefs in the usual stochastic order sense over models are positive (negative) shocks.
\thmm

\section{Analysis of Examples}

In this section, we show the applicability of our results to Examples 2 and 3. Example 2 has an explicit analytical solution and is more of an illustrative example of the structure employed in this paper. However, Example 3,
like most problems in dynamic programming,  does not have a closed-form solution and, therefore, is most amenable to analysis with our framework.

\begin{customex}{2 (contd.)}[Dynamic effort with unknown ability (\cite{ep21a})]\label{dem} 

\nf We now verify our  assumptions. The state space $\mathbb S=\{0,1\}$ and action space $\mathbb X=\{H,L\}$ are lattices. Further, the utility function is increasing in the state variable $s_{t+1}$ and for any given cost of effort $c,$ it is also supermodular in state and actions owing to its linearity. Therefore,  Assumption \ref{assumption1} is satisfied. To check for Assumption \ref{assumption2},  the expectation of any increasing function with respect to model transition functions should be increasing and supermodular in states and actions. This is again satisified given the structure of the transition functions since, irrespective of the state, higher action lead to success with probability 1 in the next period, and given the action, the probability of success is independent of the current state. Since the weighted KL-divergence is strictly concave, for every given $m>0$ we have a unique minimizer, and hence, it is identified and satisfies Assumption \ref{assumption3}. Therefore, a Berk-Nash equilibrium exists for this SMDP from Theorem 1. Notice that Assumption \ref{assumption4} is satisfied since the mapping $\theta_{Q}(m_{s})$ is increasing in $m_{s}.$ Therefore, our results in Theorem 2 are applicable for this setting. 
 We  focus on the case with unique solution and for this example, the misspecified policy function is independent of the current state. That is,
\begin{equation*}
g(s,\theta,p)   = \begin{cases}
 H, & \text{$\theta<1-c,$} \\
L, & \text{$\theta\geq 1-c.$}
\end{cases} 
\end{equation*}

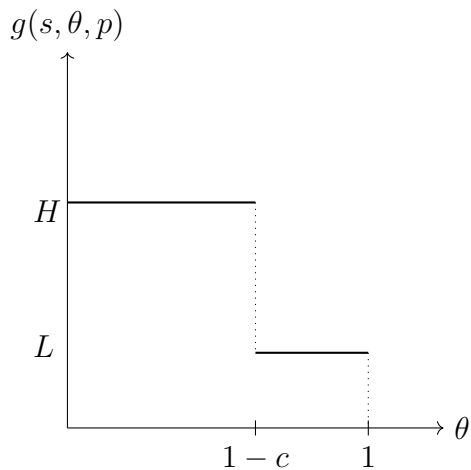
\begin{figure}[h]
\centering
 \begin{tikzpicture}
    \draw[->] (0, 0) -- (5, 0) node[right] {$\theta$};
    \draw[->] (0, 0) -- (0, 5) node[above] {$g(s, \theta, p)$};
    \draw (2.5, 0.1) -- (2.5, -0.1) node[below] {$1-c$};
    \draw[black, thick, domain=0:2.5, samples=100] plot (\x, 3);
    
    \draw[black, thick, domain=2.5:4, samples=100] plot (\x, 1);

\draw (4, 0.1) -- (4, -0.1) node[below] {$1$};
\draw[dotted] (2.5, 3) -- (2.5, 1);
\draw[dotted] (4, 1) -- (4, 0);
 \node[black, above right] at (-0.6, 2.6) {$H$};
  \node[black, above right] at (-0.6, 0.8) {$L$};
\end{tikzpicture}
\caption{An increase in the cost $c$ of high effort is a negative shock for the optimal policy correspondence, which in this case is a function.}
\end{figure}

 We now verify the prediction of our results vis-\`a-vis the analytical solution. A fall in the cost of effort $c$ is a positive shock, and therefore, leads to a higher inference of ability, $\theta^{*}=1-c,$ in the Berk-Nash equilibrium, as predicted in Theorem 2. Similarly, a fall in the cost of the effort leads to an increase in the probability of success at the steady state $m_{s}^{*}(1);$ since an increase in the  model parameter is a positive shock. Here,  the stationary distribution $m^{*}_{s}$ is increasing in the usual (first) stochastic order dominance in the cost of effort.

\end{customex}

\begin{customex}{3 (contd.)}[Savings with misperceived wealth process (EP, ADGK)] \nf


\nf First, we verify our assumptions. The state, action and parameter spaces are lattices and the utility function is increasing in the state variables, $y$ and $z$. Since the payoff function is concave in $y$ and $x,$ and $\dfrac{\dee^{2}u(y,z,x)}{\dee x\dee y}>0,$  it is  supermodular, and hence, satisfies Assumption \ref{assumption1}. Further, the model distributions  are Gaussian with mean $\alpha+\beta \ln x,$ and unit variance and therefore, satisfy Assumption \ref{assumption2}. This follows from the fact that an increase in action $x,$ increases the mean, and shifts the model distributions rightward, in the  usual order stochastic dominant sense. Assumption \ref{assumption3} is met since the Gaussian distribution is strictly log-concave, and therefore, ensures unique identification. Hence, from Theorem 1, the Berk-Nash equilibrium exists. Furthermore,  Assumption 4 is satisfied since this follows from a routine verification of the sufficient condition 1 for increasing best-fit models, under the usual stochastic order.\fn{The expression $\mathbb E_{Q(\cdot|s,x)} L(s'|s,x)$ is increasing in $(s,x)$ under the Gaussian distributions, $Q$ and $Q_{\theta}.$ For $(\alpha_{1},\beta_{1})<(\alpha_{2},\beta_{2}),$ $\mathbb E_{Q(\cdot|s,x)} L(s'|s,x)= \dfrac{1}{2\sigma^{2}}[(\alpha_{2}-\alpha_{1})(\beta_{2}-\beta_{1})(\gamma^{*}z)x]$ which is increasing in $x$ and $z$.}

Therefore, from Theorem 2, we have a clear prediction about the steady-state inferred model and the Berk-Nash equilibrium. For example, an increase in the discount factor, a positive shock to the primitives, will lead to a higher inferred return $\beta^{m}$ in the equilibrium. Similarly, an increase in the expected preference shock, another instance of a positive shock, will also lead to a higher inferred return. Moreover, since increases in beliefs about the return are a positive shock, it will perpetuate into greater savings in the steady state, and a usual order stochastic increase in the stationary wealth distribution. One advantage of our comparative statics framework is that we generate predictions about the comparative statics of the equilibria even when a closed-form solution may not exist for the equilibrium objects, as is true for this case. Further, since Assumption 5 holds, the consequent implications also hold for Theorem 4.

\end{customex}

\section{Welfare Comparisons}

We now turn towards comparing the \textit{objective} welfare of an agent facing a regular SMDP, under correctly specified and misspecified Bayesian learning and characterize an upper bound on the difference between the two instances, the costs of misspecification, in terms of the primitives of the environment. For the purposes of this section, we shall assume that the utility function is bounded and strictly concave.\fn{Another simplification we use here is working with payoff functions that have the current state and action as arguments.} Further, we shall assume that $u: \mathbb S\times \mathbb X \rightarrow \mathbb R$ is continuously differentiable in actions.

Let $g(s, \bar\theta),$  parameterized by a parameter $\bar\theta,$ denote an optimal policy function. Suppose it is well-defined under both correctly specified and misspecified environments, where $\bar\theta$ take values $\theta^{*}$ and $\theta_{*}$ under correctly specified and misspecified learning, respectively. We are interested in comparing the welfare of a correctly specified agent with parameter $\theta^{*}$ to the welfare of an agent who settles at the parameter $\theta_{*}$ asymptotically, under misspecified Bayesian learning as in Equation (1). Then, the agent's welfare under the parameter $\bar\theta,$    is the \textit{objective} ex-ante expected discounted payoff, $W(s,\bar\theta)$ of choosing the optimal action, $g(s, \bar\theta),$ and is given as, 
\begin{equation}
\label{welfarecorrect}
W(s, \bar\theta)= \mathbb{E}_{Q (\cdot|s, g(s,\bar\theta) )}\Bigg[\sum_{t=0}^{\infty} \beta^{t} u(s_{t}, g(s_{t}, \bar\theta))\Bigg], \ t=0,1,2, \ldots. 
\end{equation}

\nt Notice that for objective welfare, the expectation is solely with respect to the true transition function, $Q(\cdot|s, g(s, \bar\theta)).$ In Equation (\ref{welfarecorrect}), the form of environment affects welfare through the optimal policy action via two channels: first, through the per-period utility function, and second, through the true transition function, $Q.$ The optimal policy action $g(s, \theta_{*})$ under misspecified Bayesian learning  can theoretically be  computed in accordance with  Definition \ref{berknash} and the  \textit{subjective} distribution $Q_{\theta_{*}}.$ Our next theorem compares the welfare ranking between the two instances where our choice of metric is the one induced by sup-norm in the space of functions, $W:\mathbb S\times \Theta \rightarrow \mathbb R.$\fn{The metric induced by sup-norm is the following:  $\vert \vert W(s, \theta^{*})-W(s, \theta_{*})\vert \vert =\sup_{s}\vert W(s, \theta^{*})-W(s, \theta_{*})\vert.$ Similarly, for the policy function   $\vert \vert g(s, \theta^{*})-g(s, \theta_{*})\vert \vert =\sup_{s}\vert g(s, \theta^{*})-g(s, \theta_{*})\vert.$}

\thm Welfare under correctly specified learning $W(s,\theta^{*})$ is weakly greater than welfare under misspecified learning, $W(s,\theta_{*}).$ Further, if $m_{0}$ and $m_{1}$ denote the absolute upper bound on the utility function and the  marginal utility function, respectively, and if $||g(s,\theta^{*})-g(s,\theta_{*})||\leq \gamma,$ then 
\eq \vert \vert W(s, \theta^{*})-W(s, \theta_{*})\vert \vert\leq \dfrac{2\beta m_{0} (1-e^{-k^{*}})+m_{1}\gamma}{1-\beta},
\label{welfare}\eqq
where $k^{*}$ is the upper bound on the KL divergence of $Q (\cdot|s, g(s, \theta_{*}))$ with respect to $Q (\cdot|s, g(s, \theta^{*})).$
\thmm

We interpret the difference $\gamma$ between the two policy functions in the sup-norm as the approximation error in the space of policy functions. Then,
for a given approximation error $\gamma,$   Equation (\ref{welfare}) supplies an upper bound on the welfare comparison in terms of the primitives, namely the discount factor, and the absolute bounds on the utility, the marginal utility function, and the KL divergence. Notice that if $\gamma=0,$ that is, the policy functions under the two instances of learning are identical, then the two welfare quantities are equal, by construction.  The upper bound in Equation (\ref{welfare}) gives intuitive comparative statics, vis-\`a-vis the model primitives. For instance, the discrepancy in welfare is larger the greater the approximation error, $\gamma.$  Similarly, for a given positive approximation error, the discrepancy increases in the discount factor. That is, as the agent gets increasingly patient--reflected in a higher discount factor $\beta$--his approximation error accumulates over the horizon, leading to a greater discrepancy in welfare. Furthermore, higher bounds on the utility function and its corresponding marginal utility function lead to a greater discrepancy in the welfare. The marginal utility function plays a role in the discrepancy by attaching itself to the approximation error. Intuitively, getting the policy wrong is more pronounced if it matters more at the margins. Notice that for the upper bound on the KL divergence, we only compare the implied distributions by the true transition function $Q$ under the two policy functions.  This is because our interest lies in comparing the objective welfare under $Q$.

It is worth mentioning here that bound in Equation (\ref{welfare}) is not specific to a misspecified MDP, rather, it holds for any MDP environment with a endogenous state evolution process, when plugged with alternate policy functions. In this sense, it is related to Lemma 3.1 in \cite{sa00} which does a similar exercise in  bounding welfare discrepancies. However,  the state process in there is exogeneous. Furthermore, \cite{sa00} uses Euler residuals to provide upper bounds on the approximation error in terms of the primitives for both the policy and the value functions, which is useful for computing numerical approximations. Often, MDP environments do not have closed form solutions, thus making economists rely on numerical techniques to solve for them computationally. We hope that Theorem 5 is instructive in this regard and helps initiate conversations on the numerical approximation of Berk-Nash equilibria.

\section{Concluding Remarks and Extensions}

Models, by their very nature, offer simplified abstractions of reality, inevitably omitting certain nuances. This paper outlines conditions under which an important qualitative property, the comparative statics of decision-making and the corresponding inference, is preserved under model misspecification. The main contribution of this paper lies in establishing monotone comparative statics results for misspecified dynamic optimization problems, by a novel application of techniques in the fixed points literature (\cite{sm71}) and first used in the context of large dynamic economies (\cite{aj15}). We also illustrate the utility of these results for general interest environments. Further, we provide an upper bound on the welfare comparison between correct and misspecified learning, in terms of the primitives. We conclude by outlining several promising directions for extending our findings to more broader environments.

\medskip

\nt\textbf{Multi-dimensional parameter spaces.} Since most of the current applications in the literature predominantly involve one-dimensional models,\fn{This feature of the current literature is also noted in \cite{ep21b}.} the results presented here focus on the model (parameter) set as a compact subset of the real line. However, they could be extended to multi-dimensional parameter spaces. This would involve a suitable rehabilitation of Assumptions 3 and 4 for lattice and non-lattice multi-dimensional parameter spaces. Although one limitation of these results using the current techniques is that the comparative statics results imply all inferred parameters in equilibrium must be monotonic together, which may not be ideal for specific economic applications.

\medskip
\nt\textbf{Other forms of updating.} While our results primarily address the asymptotic behavior of Bayesian learners, it's important to note that other inference processes, such as Maximum Likelihood Estimation (MLE) and moment-based learning, can lead to the same asymptotic inferences of best-fit models. \cite{cs23} provide an illustrative example that demonstrates the identical nature of inference between Bayesian inference and methods based on likelihoods.

\medskip

\nt\textbf{Static environments and dynamic concerns.} Our results are tailored for dynamic MDP environments but they potentially are also applicable, subject to modifications, to static environments, such as the one in \cite{ep16ecma} for static games. This also spills over to settings where agents are endogenously concerned about their misspecified models, such as those in \cite{la22}.\fn{Corollary 1 in \cite{la22} presents an interesting comparative statics application to monetary policy cycles; see it for more details.} 

\nt\textbf{Monotone comparative dynamics.} The scope of our paper is limited to analyzing the comparative statics properties of the equilibrium objects in the steady state. An interesting  question arises regarding how the dynamics of the misspecified learning process might react to changes in its primitives. Specifically, we do not know how the variations in the frequency of state-action pairs, as well as the sequence of posteriors as described in Equation (1), would be influenced  because of changes in the economic primitives.  There is added complexity in the dynamics due to potential complementarities between current actions and the inference process via the role of experimentation. In this regard, the tools developed in \cite{bdrw22} for monotone comparative dynamics for stochastic games could be useful for further exploration.

\medskip
\nt\textbf{Misspecification in large economies.}  Our results are also potentially applicable to misspecified dynamic economies with a continuum of agents, such as the one conceived in \cite{mo19}. Given that the techniques we rely on (\cite{aj15}) were developed in the context of economies with a continuum of agents, the results in this paper should potentially be applicable on equilibrium concepts that \cite{mo19} develops for the boundedly rational agents working with misspecified models in macroeconomic environments. 
\newpage

\appendix

\section{Mathematical Preliminaries}
\label{mathpreliminaries}
In this section, we provide the necessary mathematical preliminaries required to go through the proofs. Let $X$ be a set. A subset $\succsim$ of $X\times X$ denotes a {\it binary relation} on $X.$ A binary relation $\succsim$ is a partial order if it is reflexive, transitive, and anti-symmetric. A partially ordered set, or a \textit{poset}, is a pair $(X, \succsim)$ that consists of a set $X$ and a partial order $\succsim$. A binary relation $\succsim$ is closed if the graph of 
$\succsim$ is a closed subset of $X \times X.$ The poset $(X, \succsim)$ is a lattice if for any $x, x^{\prime} \in X,$ the greatest lower bound (infimum) $x \wedge x^{\prime}$ and the least upper bound (supremum) $x \vee x^{\prime}$ are in $X,$ where $\wedge$ and $\vee$ denote the meet and join operations, respectively. A subset $A\subseteq X$ is a sublattice of lattice $X$ if  $A$ is a lattice that contains the meet and join (computed in $X$) for every pair of elements in $A.$  For any subset $A$ of a poset $X$, we denote the supremum and infimum of $A$ by $\sup A$ and $\inf A$, respectively. That is, $\sup A$ is the least element in $X$ such that $\sup A \succsim a$, for all $a \in A$. Similarly, $\inf A$ is the greatest element in $X$ such that $a \succsim \inf A$, for all $a \in A$. A lattice $X$ is complete if both $\inf A$ and $\sup A$ are in $X$ for any $A \subseteq X$.  A chain is a totally ordered poset. A poset $X$ is (countably) lower chain complete if any (countable) chain $A \subseteq X$ has its infimum in $X$. The poset is (countably) upper chain complete if any such chain has its supremum in $X$. The poset is (countably) chain complete if it is both upper and lower (countably) chain complete.  Let $X$ and $Y$ be subsets of $\mathbb{R}$. Set $Y$ dominates $X$ in the strong set order if for any $x$ in $X$ and $y$ in $Y$, we have $\max \left\{x,y\right\}$ in $Y$ and $\min \left\{x, y\right\}$ in $X$.

Let $\mathcal M(X)$ denote the space of probability measures defined on a compact subset of $X \displaystyle\subset \mathbb{R}^n$. Although even if $X$ is a lattice, the poset $(X,\succsim_{st})$ is not a lattice as pointed in \cite{kko77}. For e.g., for $(\mathbb R^{2},\succsim_{st})$ let $p_{1}= 0.5 (\epsilon_{a}+\epsilon_{b}), p_{2}= 0.5 (\epsilon_{a}+\epsilon_{c}),p_{3}= 0.5 (\epsilon_{c}+\epsilon_{b}),p_{4}= 0.5 (\epsilon_{a}+\epsilon_{d}),$ where $a=(0,0), b=(0,1), c=(1,0), d=(1,1).$ Then, both $p_{3}$ and $p_{4}$ are supremum, which is a contradiction. However, $(\mathbb R^{2},\succsim_{st})$ is chain-complete.\fn{This also holds true for increasing and convex order, $\succsim_{icx}$.} A correspondence $T: X\rightarrow 2^{Y} $ is upper-hemicontinuous at a point $x_{0}\in X$ if for any sequence $\{x_{n}\}_{n\in \mathbb N}$ such that $\{x_{n}\}\rightarrow x_{0}, \ y_{n}\in T(x_{n}), y_{0}\in T(x_{0})$ implies $y_{n}\rightarrow y_{0}.$  We will require the following existence theorem in \cite{sm71} for chain-complete (non-lattice) spaces, and refer the interested reader to his paper for further details.

\begin{theorem*}[Smithson's fixed point theorem (1971)] 
\label{thm-sm71}
Let $X$ be a chain-complete poset equipped with partial order $\succsim,$ and $T:X\rightarrow 2^{X}$ a Type I (Type II) monotone correspondence. Suppose for any chain $C$ in $X$ and any monotone selection $f$ from the restriction of $T$ to $C,$ $f: C\rightarrow X, $ there exists $y_{0}\in T(\sup C)$ such that $ y_{0} \succsim f(x)$ for all $x\in C.$ Then if there exists a point $e \in X$ and a point $y\in T(e)$ such that $y\succsim e,$ then $T$ has a fixed point.
\end{theorem*}

\section{Proofs}

This section is divided into four parts. Part (i) proves two auxiliary lemmas that are instrumental for our main results. Part (ii) proves Theorems 1  and 2 of the paper and part (iii) proves Theorems 3 and 4, respectively. Part (iv) does the welfare comparison.

The proofs for Theorems 1 and 2 are structured in three steps, with the proof technique similar to that in \cite{aj15}.   For Theorem 2, the \textit{first} step involves showing that for any fixed model distribution $\mu\in \mathcal M_{1}(\Theta)$, the set of stationary distributions on states and actions, $m^{*}$, induced by the optimal policy correspondence $G,$ will be Type I (Type II) monotonic in the primitives $p$. The \textit{second} step involves constructing a mapping $\hat\theta$ that for each given $\mu$ and $p$ yields a set of model distributions, $\mu's$. It is the fixed points of this map that are the equilibrium model distributions $\mu^{*},$ given $p$. Finally, the \textit{third }step involves the least and greatest selections from this map will be increasing in $p.$ This in turn leads us to give a new existence proof of Theorem 1 that relies on the monotonicity and identification properties of the equilibrium map, $T$. The rest of the proofs (Theorems 3 and 4) shall follow an analogous structure. Proof of Theorem 5 is based on the Taylor expansion of the utility function and on an application of a new result proven by \cite{ca22} on entropy bounds.

The equilibrium mapping $T$ associated with the Berk-Nash equilibrium (Definition 5)  is a set-valued function on the product space of probability measure on states and actions, and parameter space, $T: W\rightarrow 2^W,$ where $W=\mathcal M_{1}(\mathbb S\times \mathbb X)\times \mathcal M_{1}(\Theta)$ such that $\mathcal{T}(m, \mu)=\mathcal{M}(m, \mu) \times \mathcal{M}\left(\Theta_Q(m)\right)$, where
\begin{equation}
\label{adjoint}
(m, \mu) \mapsto \mathcal{M}(m, \mu) \equiv\left\{m^{\prime} \in \mathcal M_{1}(\mathbb S\times \mathbb X): m^{\prime} \in \mathcal{F}(\mu) \text { \& } m_{\mathbb{S}}^{\prime}(\cdot)=\int_{\mathbb S\times \mathbb X}Q(\cdot|s,x)m(\dee s, \dee x)\right\}
\end{equation}
for any $\mu \in \mathcal M_{1}(\Theta), \mathcal{F}(\mu)$ is the set of all $m^{\prime}$ that satisfy the conditions of optimality and  stationarity as defined in Definition 5.

\smallskip
\nt{(i)\textbf{  Auxiliary Lemmas}}

\nt Towards this end, we begin by proving two lemmas. They adapt lemmas in \cite{aj15} that are established for exogenous shock processes to our  setting of endogenous Markov decision process.  Lemma \ref{policyselection}  shows that under assumptions 1 and 2, the optimal policy correspondence, defined in Eq (\ref{stationarypolicy}), $G:\mathbb S\times P\rightarrow 2^{X}$ is increasing in the strong-set order and the least and the greatest selection are increasing in $s.$
\lm 
\label{policyselection}
Let the regular SMDP satisfy assumptions 1 \& 2. Then for any given primitive $p,$ the optimal policy correspondence $G: \mathbb S\times P \rightarrow 2^\mathbb X$ is increasing in the state $s$ in the strong set order. In particular, it has a least and a greatest selection, and they  are increasing in state $s$.\fn{We suppress the dependence on  $\mathcal M_{1}(\Theta)$ for notational convenience. Theorem 3.9.2 in \cite{to98} establishes conditions under which finite and infinite period discounted MDPs have increasing policy selection.}
\prf
Given a model distribution $\mu,$ and corresponding to a Bellman solution $V$ of Eq. (\ref{berknash}),
\eq
 G(s,\mu, p)=\argmax_{x\in \mathbb  X}\displaystyle \ \int_{S} \{u(s,x, s')+ \delta \ V(s',\mu, p)\} \bar{Q_{\mu}}(\dee s'|s, x),
\eqq
where  $\bar Q_{\mu}=\displaystyle\int_{\Theta}Q_{\theta}\mu(d\theta).$  Given Assumptions \ref{assumption1} and \ref{assumption2}, and by the application of Theorem 3.9.2 (page 165, \cite{to98}), the expression on the right-hand side is supermodular in $(s,x)$. Therefore, the optimal policy correspondence will be increasing in $s$ in the strong-set order. Further, given the optimal correspondence, the greatest and least selection exist and are increasing in the state. \prff
\lmm

\nt The next lemma transfers the monotonic nature of the optimal policy correspondence to its corresponding fixed point map, $T,$ in terms of Type I (Type II) monotonicity.

 \lm  
 \label{adjointmonotonicity}
 If the optimal policy correspondence $G: \mathbb S\rightarrow 2^\mathbb X$ has an increasing greatest (least) selection, then the fixed point correspondence $T$ is Type I (Type II) monotone with respect to $\succsim_{st}$.  Further, if $G$ depends on a primitive $p \in P$ such that $G: \mathbb S \times P \rightarrow 2^\mathbb X$ and the greatest (least) selection from $G$ is increasing in $p$, then the fixed point correspondence $T$ indexed by $p$, $T_p$ is Type I (Type II) monotone in $\mathcal M_{1}(\mathbb S\times \mathbb X) $ with respect to $\succsim_{st}$.
 \label{adjmonotone}
\lmm

\prf We only prove the above statement for the greatest selection in the Type I case.\fn{The rest of the cases involving least selection and Type II monotonicity follow analogously.} Consider probability measures, $\nu_1, \nu_2 \in \mathcal{M}_{1}(\mathbb S\times \mathbb  X)$ such that $\nu_{2}\succsim_{st} \nu_{1}.$ To prove that the fixed point correspondence $T$ is Type I monotone, we need to  show that for any $\lambda_1 \in$ $T\nu_1$, there exists a $\lambda_2 \in T\nu_2$ such that $\lambda_2 \succsim_{st} \lambda_1.$ That is, if $\lambda_1 \in T\nu_1,$ then there exists a measurable selection $g_1:\mathbb S\rightarrow \mathbb X,$ such that, for all $A \times B \in \mathcal{B}(\mathbb S\times \mathbb X)$,
$$\lambda_1(A, B)=\int_{\mathbb S \times \mathbb X} Q(A|s, g_1(s)) \chi_B\left(g_1(s)\right) \nu_1(\dee s, \dee x)$$
and therefore, there must exist a  selection $g_2 $ such that  $\lambda_2 \in T\nu_2$, and $\lambda_{2}\succsim_{st} \lambda_{1}.$
From Lemma 1, we know that such a greatest selection exists, and therefore, for all $A \times B \in \mathcal{B}(\mathbb S\times \mathbb X),$
\begin{flalign*}
\lambda_{2}\succsim_{st}\lambda_{1} \iff \int_{\mathbb S} f(s, g_2(s)) Q(\dee s| s, g_2(s)) \nu_2(s)&  \geq \int_{\mathbb S} f(s, g_2(s)) Q(\dee s| s, g_2(s)) \nu_1(s)\\
& \geq  \int_{\mathbb S} f(s, g_1(s)) Q(\dee s| s, g_1(s))\nu_1(s)
\end{flalign*}

The first inequality follows from $\nu_{2}\succsim_{st} \nu_{1}.$ The second inequality follows from $Q$ being monotone and $g_{2}$ being the greatest selection. With an identical argument, one can prove for the case where the greatest selection from $G$ is to be increasing in primitive $p,$ and therefore, $T$ indexed by $p, T_{p}$ is Type I (Type II) monotone in $\mathcal M_{1}(\mathbb S\times \mathbb X),$ with respect to $\succsim_{st}.$ 
\prff

\nt We are now ready to prove Theorems 1 and \ref{thmpositive}. Throughout the proof, and without loss of generality, primitive $p$ is restricted to the set $P \equiv\{p_{1}, p_{2}\}$ ordered by $p_{2}\succsim p_{1},$ where the ordering depends on the primitive being considered.

\medskip

\nt (ii)
\nt{\underline{\textbf{Proofs for Theorems 1 and 2:}}}
The \textit{first} step involves showing for a given model distribution $\mu \in \mathcal M_{1}(\Theta),$ the set of Berk-Nash equilibriua $m$ induced by the optimal policy correspondence $G$ will be Type I (Type II) increasing in the primitives $p\in P.$ To this end, we first show that under a positive shock, the fixed point correspondence $T$ is Type I (Type II) increasing in the primitives $p.$ The fixed points of this correspondence are the Berk-Nash equilibrium $m$ for a given model distribution $\mu,$  and finally using a result in \cite{aj15}, we show that the set of Berk-Nash equilibriua for a given model distribution is Type I (Type II) increasing in the primitives.

 Because of Lemma \ref{policyselection}, the stationary optimal policy correspondence $G$ will have a least and a greatest selection that will be increasing  in $s$ and therefore, by Lemma \ref{adjmonotone},  for a given model distribution $\mu,$ and primitive $p,$  $T_{\mu, p}: \mathcal{M}_{1}\left(\mathbb S\times \mathbb X\right) \rightarrow 2^{\mathcal{M}_{1}\left(\mathbb S\times \mathbb X\right)}$ induced by the stationary optimal policy correspondence, $G,$  defined in (\ref{stationarypolicy}) is Type I (Type II) monotone with respect to $\succsim_{st}$.  Now from a routine modification of Theorem B3 in \cite{aj15},\fn{Theorem B3 in \cite{aj15} builts on the existence theorem in \cite{sm71}. It states the following: Assume that the equilibrium mapping $T$ is either Type I or type II monotone. In addition, assume that the set of measures on state and actions has an infimum. Then $T$ has fixed point. In addition, the fixed-point correspondence is upper hemicontinuous if $T$ is upper hemicontinuous.} the set of fixed points, $F: \mu \times P \rightarrow 2^{\mathcal{M}_{1}\left(\mathbb S\times \mathbb X\right)}$, given by $F\left(\mu, p\right)=\left\{m \in \mathcal{M}_{1}(\mathbb S\times \mathbb X): m \in T_{\mu, p} m\right\}$ is non-empty valued and upper hemicontinuous. The proof of upper-hemicontinuity of $T_{\mu,p}$ follows for the finite case follows from Claim B (page 744) in EP, while for the infinite case, specifically compact Euclidean spaces here, it follows by invoking the conditions in Definition 3 on a regular SMDP, as proved in \cite{adgk22}.  From Lemma 2, $T_{\mu, p}$ is  Type I (Type II) monotone in $p$, and therefore, from the monotonicity theorem of \cite{aj15} in the main text (page 15), the set of fixed points $F$ will be non-empty and Type I (Type II) monotone in $p.$ Hence, the set of Berk-Nash equilibriua $m$ induced by the optimal policy correspondence $G$ will be Type I (Type II) increasing in the primitives $p$. This completes the first step.
\bigskip

\nt The \textit{second} step constructs a mapping $\hat\theta:\mathcal M_{1}(\Theta)\times P\rightarrow 2^{\mathcal M_{1}(\Theta)}$ that for each given model distribution $\mu$ and primitive $p\in P,$ yields a set of best-fit model distributions,
\begin{equation}
\hat{\theta}(\mu, p)=\{\theta(m)\equiv \mathcal \argmin_{\theta\in\Theta} K_{Q}(m,\theta): m \in F(\mu, p)\}.
\end{equation}
\nt  A  model distribution $\mu^{*}$ is an equilibrium belief  if and only if $\mu^* \in \hat{\theta}(\mu^{*}, p)$. By assumption 3 of point identification  and therefore, by uniqueness, for each $m,$ there will be a non-empty unique $\theta(m).$ That is, one has a Dirac measure on $\theta(m).$ From Berge's Maximum Theorem, $\theta$ is continuous\fn{The function $\theta_{Q}(m)$ is continuous if it is continuous in the weak$*$topology on its domain.}  and given the upper hemicontinuity of $F\left(\mu, p\right)$, $\hat{\theta}$ will  be upper hemi-continuous.  The rest of this step follows \cite{aj15}. For a fixed $\mu,$ and given $F(\mu, \cdot)$ is Type I and Type II monotone in $p,$ and increasing $\theta(m)$ under Assumption 4, one can use their Theorem 4 (pp. 601) to conclude that the least and greatest selections from $\hat{\theta}(\mu, \cdot)$  will be increasing in $p$ holding $\mu$ fixed.\fn{Theorem 4 in \cite{aj15} guarantees least and greatest selections from a fixed point correspondence. It may be possible to dispense with identification (Assumption 3) if we were to use more general divergence measures apart from the Kullback-Leibler divergence. However, given the criticality of the KL divergence, uniqueness is a feature we require. It is also something that the current literature has emphasized in terms of applications; see \cite{ep21a, ep21b}.}\\

\nt The \textit{third} and \textit{final} step shows that if the least and the greatest selections of the upper-hemicontinuous fixed
point correspondence $\hat{\theta}$ are increasing in $p$, then the fixed points are increasing in $p$. Since under Assumption 4, $\theta$ is a monotonic function of $m,$ therefore, $\mu_{\min } \equiv \theta(\delta_{\inf \mathbb S\times \mathbb X}) $ and $\mu_{\max} \equiv \theta(\delta_{\sup \mathbb S\times \mathbb X})$, where $\delta_{\mathbb S\times \mathbb X}$ denotes the degenerate measure on  $\mathbb S\times \mathbb X$ with its mass at $(s,x)$. Therefore, $\mu \succsim \mu_{\text {min }}$ for all $\mu \in \hat{\theta}\left(\mu_{\min }\right)$ and $\mu \precsim \mu_{\max }$ for all $\mu \in \hat{\theta}\left(\mu_{\max }\right)$. Therefore, for every $p \in P$, $\hat{\theta}(\cdot, p):\left[\mu_{\min }, \mu_{\max }\right] \rightarrow 2^{\left[\mu_{\min }, \mu_{\max }\right]}$. Notice in step 2 that $F(\mu,p)$ is a convex-valued set of fixed points since $T_{\mu,p}$ is convex-valued; the proof of convex-valuedness follows from EP. Therefore, the set of fixed points from $F$ is convex-valued and given $\theta(m)$ is a continuous function, $\hat\theta(\mu,p)$ is convex-valued. Hence, from \cite{aj13},  $\hat{\theta}(\cdot, p):\left[\mu_{\min }, \mu_{\max }\right] \rightarrow 2^{\left[\mu_{\min }, \mu_{\max }\right]}$ is upper hemicontinuous and convex valued and for each fixed value of $\mu \in[\mu_{\min},\mu_{\max}],$  has least and greatest selections and  are increasing in $p$ and from  Corollary 2 in \cite{mr94} the least and greatest fixed points $\mu^{*} \in \hat{\theta}(\mu^{*}, p)$ will be increasing in $p$. Hence, we have proven that the least and greatest inferred models are increasing in the primitives under a positive shock to a regular SMDP. Furthermore, From Theorem 2.8.3 in \cite{to98}, and by treating a change in the beliefs $\mu$ in the usual stochastic order sense as a  change in primitive as above, the remaining part of Theorem 2 follows. 

 Existence is yielded in Step 2 by the Kakutani-Fan-Glicksberg Theorem since our map $\hat\theta(\mu,p)$ is convex-valued, upper hemi-continuous, and as is shown in EP, $\mathcal M_{1}(\Theta)$ are locally convex Hausdorff spaces.

\qed

\medskip 
\nt(iii) {\underline{\textbf{Proof for Theorem 3:}}}  The parameter space $\Theta\subseteq \mathbb R$ is one-dimensional, and hence, the weighted KL divergence satisfies quasimodularity by triviality on the parameter space, $\Theta\subseteq \mathbb R$.  Hence, from \cite{to78} and \cite{ms94}, given an increase in the parameter space from $\Theta_{1}$ to $\Theta_{2}$ in the strong-set order, the set of minimizers are increasing in the strong-set order, i.e.,
$$\Theta(m;\Theta_{1})\equiv\argmin_{\theta\in \Theta_{1}} K_{Q}(m,\theta) \subseteq \argmin_{\theta\in \Theta_{2}} K_{Q}(m,\theta)\equiv\Theta(m;\Theta_{2})$$

\nt Therefore, if a change in model distribution is a positive shock then the proof follows on the lines of Theorem 2. Thus, this completes the proof. \qed
\medskip

\nt{\underline{\textbf{Proof for Theorem 4:}}} Consider probability measures, $\nu_1, \nu_2 \in \mathcal{M}(\mathbb S\times \mathbb  X)$ such that $\nu_{2}\succsim_{icx} \nu_{1}.$  The proof of Type I and Type II monotonicity is similar to that of Theorems 1 and 2, albeit requires that we work with increasing convex orders on the set of states and actions.

To prove that the fixed point correspondence $T$ is Type I monotone in $\succsim_{icx}$, we need to  show that for any $\lambda_1 \in$ $T\nu_1$, there exists a $\lambda_2 \in T\nu_2$ such that $\lambda_2 \succsim_{icx} \lambda_1,$ that is, if $\lambda_1 \in T\nu_1,$ then there exists a measurable selection $g_1:\mathbb S\rightarrow \mathbb X,$ such that, for all $A \times B \in \mathcal{B}(\mathbb S\times \mathbb X)$,
$$\lambda_1(A, B)=\int_{\mathbb S \times \mathbb X} Q(A|s, g_1(s)) \chi_B\left(g_1(s)\right) \nu_1(\dee s, \dee x)$$
and therefore, there must exist a  selection $g_2 $ such that  $\lambda_2 \in T\nu_2$, and $\lambda_{2}\succsim_{icx} \lambda_{1}.$
From Lemma 1, we know that such a greatest selection exists, and therefore, for all $A \times B \in \mathcal{B}(\mathbb S\times \mathbb X),$
\begin{flalign*}
\lambda_{2}\succsim_{icx}\lambda_{1} \iff \int_{\mathbb S} f(s, g_2(s)) Q(\dee s| s, g_2(s)) \nu_2(s)&  \geq \int_{\mathbb S} f(s, g_2(s)) Q(\dee s| s, g_2(s)) \nu_1(s)\\
& \geq  \int_{\mathbb S} f(s, g_1(s)) Q(\dee s| s, g_1(s))\nu_1(s)
\end{flalign*}

The first inequality follows from $\nu_{2}\succsim_{icx} \nu_{1}.$ The second inequality follows from $Q$ being monotone and $g_{2}$ being the greatest selection. The only adjustment in the proof that remains is that the best-fit set construction needs to incorporate Assumption 5 for it to have them monotonic, i.e., the \textit{second} step constructs a mapping $\hat\theta$ that for each given model distribution $\mu$ and primitive $p\in P,$ yields a set of model distributions,
\begin{equation}
\hat{\theta}(\mu, p)=\{\theta(m)\equiv \argmin_{\theta\in\Theta} K_{Q}(m,\theta): m \in F(\mu, p)\}.
\end{equation}
 
\nt Hence, this completes the proof.

\qed
\medskip 

\nt (iv) {\underline{\textbf{Proof for Theorem 5:}}} The first part follows from the optimality of $g(s, \theta^{*})$ over $g(s, \theta_{*}),$ under the  objective  welfare function $W(s, \theta^{*}),$ where the expectation is taken with respect to the true transition function, $Q.$  Next, under correct learning parameterized by $\theta^{*},$ and for an initial state $s_{0},$ the objective welfare is given as,
\begin{equation}
W(s_{0}, \theta^{*})= \mathbb{E}_{Q (\cdot|s_{t-1}, g(s_{t-1},\theta^{*}) )}\Bigg[\sum_{t=0}^{\infty} \beta^{t} u(s_{t}, g(s_{t}, \theta^{*}))\Bigg], \ t=0,1,2, \ldots. 
\end{equation}
\nt Unwrapping the above expression,
\eq
\begin{aligned}
\label{correctwelfare}
W(s_{0}, \theta^{*})= u(s_{0}, g(s_{0}, \theta^{*}))+\beta\int_{S}u(s_{1}, g(s_{1}, \theta^{*}))Q(\dee s_{1}|s_{0}, g(s_{0}, \theta^{*}))+...\\ 
\beta^{t}\int_{S}u(s_{t}, g(s_{t}, \theta^{*}))Q(\dee s_{t}|s_{t-1}, g(s_{t-1}, \theta^{*}))+...
\end{aligned}
\eqq

\nt Now, with the Berk-Nash parameter $\theta_{*}$, the corresponding objective welfare under misspecified learning is,
\eq
\begin{aligned}
\label{incorrectwelfare}
W(s_{0}, \theta_{*})= u(s_{0}, g(s_{0}, \theta_{*}))+\beta\int_{S}u(s_{1}, g(s_{1}, \theta_{*}))Q(\dee s_{1}|s_{0}, g(s_{0}, \theta_{*}))+...\\ \beta^{t}\int_{S}u(s_{t}, g(s_{t}, \theta_{*}))Q(\dee s_{t}|s_{t-1}, g(s_{t-1}, \theta_{*}))+...
\end{aligned}
\eqq

\nt Our objective is to find an upper bound for $\vert \vert W(s_{0},\theta^{*})-W(s_{0},\theta_{*}) \vert \vert,$ where ||$\cdot$|| denotes the sup-norm in the function space. Comparing the first terms in Equations (\ref{correctwelfare}) and (\ref{incorrectwelfare}) and by a Taylor expansion of $u(s_{0},g(s_{0}, \theta^{*}))$ on $u(s_{0},g(s_{0}, \theta_{*}))$, given that we assume that it is continuously differentiable and strictly concave, we have,
$$\vert\vert u(s_{0}, g(s_{0}, \theta^{*}))-u(s_{0}, g(s_{0}, \theta_{*}))\vert\vert \leq \vert \dfrac{\dee u}{\dee g}\vert_{g(s_{0}, \theta_{*})} ||g(s_{0}, \theta^{*})-g(s_{0}, \theta_{*})||. $$
\nt The inequality follows from the concavity of $u$ in policy function $g$ and from taking the sup-norm on both sides. Similarly, for the second term, we have,
\eq
\begin{aligned}
&\beta||\big(\int_{S}u(s_{1}, g(s_{1}, \theta^{*}))Q(\dee s_{1}|s_{0}, g(s_{0}, \theta^{*})) - \int_{S}u(s_{1}, g(s_{1}, \theta_{*}))Q(\dee s_{1}|s_{0}, g(s_{0}, \theta_{*}))\big)||\\
&\leq \beta\big (\int_{S}|u(s_{1}, g(s_{1}, \theta_{*}))| \cdot ||Q(\dee s_{1}|s_{0}, g(s_{0}, \theta^{*}))-Q(\dee s_{1}|s_{0}, g(s_{0}, \theta_{*}))|| + 
\vert \dfrac{\dee u}{\dee g}\vert_{g(s_{1}, \theta_{*})} ||g(s_{1}, \theta^{*})-g(s_{1}, \theta_{*})|| \big)\\
\end{aligned}
\eqq
\nt This again follows from the strict concavity of $u,$ a corresponding Taylor-expansion of the utility function $u,$ and the fact that the integral of the density function (always positive) is 1. 

While we have corresponding upper bounds for the utility function, its first derivative, and the difference in the policy functions under the two learning instances, we still need to establish a corresponding upper bound for the total-variation (TV) norm for densities $Q.$ Towards this end, we rely on a lemma by \cite{ca22}, that provides an almost stricter bound on the TV-norm between densities.\fn{Our choice of this particular bound over the well-known Pinsker bound is that as the KL divergence gets larger, the Pinsker bound becomes vacuous and exceeds the trivial bound of 1. The BH bound, however, never exceeds 1; see \cite{ca22} for further details.}

\begin{theorem*}[Bretagnolle–Huber bound, \cite{ca22}]
For any two probability distribution functions $Q_{1}, Q_{2}$ over $\mathbb R,$
$$d_{TV}(Q_{1}, Q_{2})=||Q_{1}-Q_{2}||_{1}\leq 2\sqrt{1-e^{-KL(Q_{1}||Q_{2})}},$$
where $d_{TV}$ is the total variation norm between densities $Q_{1}$ and $Q_{2},$ and $KL$ is the relative entropy between $Q_{1}$ and $Q_{2}.$
\end{theorem*}

\nt Therefore, the first part of the right-hand side expression in Eq.(23) can be bounded as,
$$ \beta\big (\int_{S}|u(s_{1}, g(s_{1}, \theta_{*}))| \cdot ||Q(\dee s_{1}|s_{0}, g(s_{0}, \theta^{*}))-Q(\dee s_{1}|s_{0}, g(s_{0}, \theta_{*}))||\big)\leq 2\beta m_{0}\sqrt{1-e^{-KL(\gamma, Q)}}. 
$$
where the KL distance, $KL_{1}(\gamma, Q),$ depends on the approximation error $\gamma,$ and the true transition function, $Q,$ and also the particular state realization. 
\nt For the second part, we have,
\eq
\begin{aligned}
&\vert \dfrac{\dee u}{\dee g}\vert_{g(s_{1}, \theta_{*})} ||g(s_{1}, \theta^{*})-g(s_{1}, \theta_{*})||\leq m_{1}\gamma.
\end{aligned}
\eqq

\nt By repeating the above step for all the terms in $W(s,\theta^{*})$ and $W(s,\theta_{*})$, we have,  
\begin{align*}||W(s,\theta^{*})-W(s,\theta_{*})||&\leq m_{1}\gamma+ \beta(m_{1}\gamma+2m_{0}\sqrt{1-e^{-KL_{1}(\gamma, Q)}})+\beta^{2}(m_{1}\gamma+2m_{0}\sqrt{1-e^{-KL_{2}(\gamma, Q)}})+\ldots\\
&\leq\dfrac{m_{1}\gamma+2m_{0}\beta\sqrt{1-e^{-KL_{*}(\gamma, Q)}}}{1-\beta} 
\end{align*}

\nt where $KL_{*}(\gamma, Q)$ is the upper bound on the sequence of KL distances, $\{KL_{1}, KL_{2}, \dots, KL_{n}...\}$.  Hence, we have established our upper bound between the two welfare quantities, and this completes the proof.
\qed

\setlength{\bibsep}{5pt}
\setstretch{1.05}
\newpage

\bibliographystyle{econometrica.bst} 
\bibliography{References.bib}

\end{document}